\documentclass[format=sigconf,nonacm=true]{acmart}
\usepackage{float}
\usepackage{graphicx}
\usepackage{supertabular}
\usepackage{enumerate}
\usepackage{calc}
\usepackage{arrayjobx} 
\usepackage{multido} 

\begin{document}

\title{Report of the Third Global Experimentation for Future Internet (GEFI 2018) Workshop}
\subtitle{October 25-26, 2018, Tokyo, Japan}
\author{Mark Berman} \affiliation{GENI Project Office, Raytheon BBN Technologies}
\author{Timur Friedman} \affiliation{Sorbonne Universit\'e}
\author{Abhimanyu Gosain} \affiliation{Northeastern University}
\author{Kate Keahey} \affiliation{University of Chicago}
\author{Rick McGeer} \affiliation{US Ignite}
\author{Ingrid Moerman} \affiliation{imec}
\author{Akihiro Nakao} \affiliation{University of Tokyo}
\author{Lucas Nussbaum} \affiliation{Universit\'e de Lorraine}
\author{Kristin Rauschenbach} \affiliation{Notchway Solutions}
\author{Violet Syrotiuk} \affiliation{Arizona State University}
\author{Malathi Veeraraghavan} \affiliation{University of Virginia}
\author{Naoaki Yamanaka} \affiliation{Keio University}

\begin{abstract}
The third Global Experimentation for Future Internet (GEFI 2018) workshop was held
October 25-26, 2018 in Tokyo, Japan, hosted by the University of Tokyo.
A total of forty-four participants attended, representing Belgium, Brazil,
China, Denmark, France, Ireland, Japan, the Republic of Korea, and the United States.
The workshop employed a mixed format of presentations and open group discussions
to advance multi-national coordination and interoperation of research infrastructure
for advanced networking and computer science research.

Major topic areas included: softwareization and virtualization of radios and networks;
testbed support for networking experiments; EdgeNet; a federated testbed of elastic optical
networks; and reproducibility in experimentation. Workshop goals included
both the formulation of specific new research collaborations and strategies for coordination
and interoperation of research testbeds.

Workshop outcomes include a variety of new and ongoing collaborative efforts.
Participants in the session on a federated testbed of elastic optical networks agreed to
pursue the development of optical ``white boxes'' in support of the creation of such elastic
optical testbeds. Key participants plan to form a research coordination network or similar
structure to organize future activity.
The reproducibility session highlighted parallels between
difficulties in reproducing experiments in computer science research and similar challenges
in other scientific disciplines. Participants discussed both training and tooling approaches
to addressing the situation.
The EdgeNet session reviewed the considerable progress achieved in EdgeNet deployment
since its conception at GEFI 2017 and discussed possibilities for growth in conjunction
with existing research testbed infrastructure.
The session on softwareization and virtualization of radios and networks explored both the
potential and challenges of software-based approaches to driving and supporting collaboration
among the diverse technical disciplines needed to bridge the radio and networking
research communities. The participants sought to identify strategies for effective use of 
open-source software and hardware platforms in future research infrastructure.
The session on networking experiments explored the current status of
testbeds, the current and upcoming requirements from experimenters, and the
plans for testbeds evolutions and new testbeds to address those new
needs.
\end{abstract}

\acmConference[GEFI2018]{3rd GEFI Workshop}{October 25-26, 2018}{Tokyo}
\setcopyright{none}
\settopmatter{printfolios=true}
\authorsaddresses{}
\acmPrice{}
\acmDOI{} 
\acmISBN{} 

\maketitle


\newcommand{\initPresentations}{
\newarray\PresentationAuthor
\newarray\PresentationTitle
\newarray\PresentationLink
\newarray\PresentationDescription
\newcounter{PresentationCount}
}

\newcommand{\formatPresentationWithFootnote}{
\PresentationAuthor(\value{PresentationCount}):
\textit{\PresentationTitle(\value{PresentationCount})}
\footnote{\PresentationLink(\value{PresentationCount})}
\newline\PresentationDescription(\value{PresentationCount})
}
\newcommand{\formatPresentationWithoutFootnote}{
\PresentationAuthor(\value{PresentationCount}):
\emph{\PresentationLink(\value{PresentationCount})}.
\newline\PresentationDescription(\value{PresentationCount})
}

\newcommand{\formatPresentation}{\formatPresentationWithoutFootnote}

\newcommand{\presentation}[4]{
\stepcounter{PresentationCount}
\PresentationAuthor(\value{PresentationCount})={#1}
\PresentationTitle(\value{PresentationCount})={#2}
\PresentationLink(\value{PresentationCount})={\href{#3}{#2}}
	\PresentationDescription(\value{PresentationCount})={#4}
	\formatPresentation
}

\newcommand{\insertPresentationSummary}{
\begin{enumerate}
	\multido{\iPresentation=1+1}{\value{PresentationCount}}{
		\item \PresentationAuthor(\iPresentation)
		\newline
		\emph{\PresentationLink(\iPresentation)}.
	}
\end{enumerate}
}


\initPresentations

\section[samepage]{Introduction and Motivation}

\subsection{GEFI Purpose}

\fbox{
\begin{minipage}{(\columnwidth-3\parindent)}
\emph{GEFI Mission Statement}

The participating research communities wish to perform collaborative research, on the basis of equality and reciprocity, in areas of mutual interest, which may be characterized as:
\begin{enumerate}[(a)]
	\item Investigations of the research infrastructures suitable for hosting at-scale experimentation in future internet architectures, services, and applications, and use of such infrastructures for experimental research.
	\item We envision that our collaboration will encompass joint specification of system interfaces, development  of interoperable systems, adoption of each other's tools, experimental linkages of our testbeds, and experimentation that spans our infrastructures.
	\item We further envision that students and young professors from participating nations will visit each other and collaborate deeply in these activities, in hopes of sparking friendships and life-long research collaborations between the communities.
\end{enumerate}
\end{minipage}
}

The Global Experimentation for Future Internet (GEFI) collaboration is an international research initiative with
a mission to encourage collaborative research across international boundaries in a set of topics that are 
mutually interesting to researchers in each participating country.

In keeping with GEFI's goals of encouraging testbed-supported research,
representatives of many research testbeds participated in the workshop.
See Table \ref{tab:testbeds} for a summary of testbeds represented.

\begin{figure*}
\includegraphics[width=\textwidth]{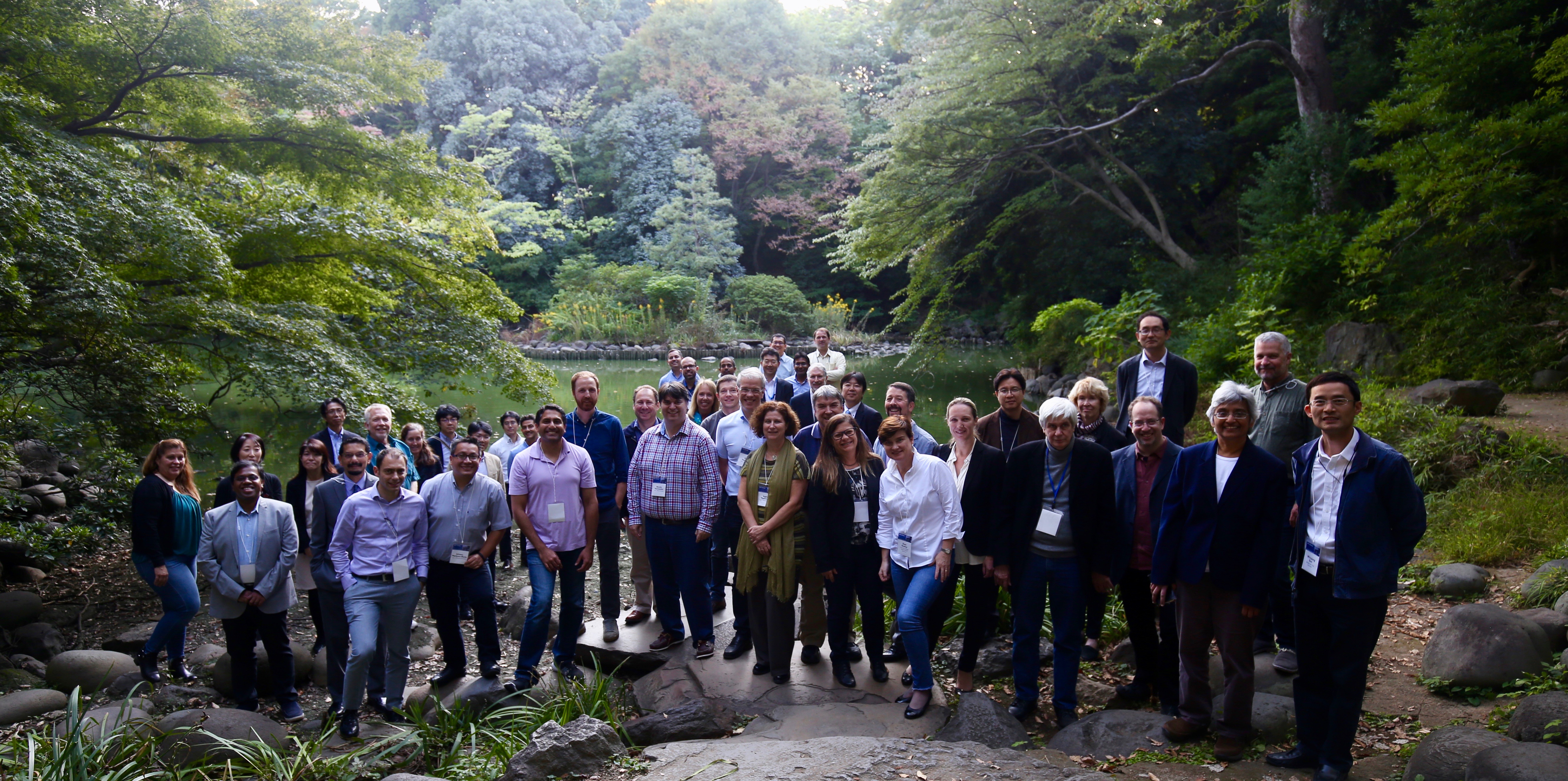}
\caption{GEFI 2018 participants gather beside Sanshiro Pond.}
\label{fig:group-photo}
\end{figure*}

\subsection{Workshop History and Predecessor Events}

This report covers the third GEFI workshop.  The first GEFI workshop was held April 18-20, 2016,
in conjunction with NetFutures 2016.\footnote{See \url{http://doc.fed4fire.eu/gefi2016} and
\url{http://netfutures2016.eu/programme/global-experimentation-for-the-future-internet/} for more information
on GEFI 2016.}
The second GEFI workshop was held October 25-26, 2017, in Rio de Janeiro, Brazil.\footnote{See
\url{http://indico.rnp.br/conferenceDisplay.py?confId=243} for more information on GEFI 2017.}

The purpose of these workshops is to create an environment for the direct exchange
of information connecting the developers and maintainers of testbeds and related research
infrastructure, as well as the researchers who make use of these capabilities.
These interactions have proven highly effective in building successful collaborations and
coordinating activities among participating research infrastructures.

The GEFI workshops represent the logical successor to the GENI-FIRE collaboration,
as well as a number of other bilateral research collaborations.
\begin{itemize}

	\begin{item}
	\textbf{GENI-FIRE Collaboration}

	Building on years of previous collaboration, the Global Environment for Network Innovations (GENI) 
	project in the US and the EU's Federation for Future Internet Research and Experimentation (Fed4FIRE)
	formalized a collaborative effort in 2013. The purpose is to encourage coordination and interoperation between
	their testbeds and research communities. In additional to numerous informal exchanges, 
	this collaboration led to four workshops.
	\begin{itemize}
		\item October 14-15, 2013 (Leuven, Belgium)
		\item May 5-6, 2014 (Cambridge, MA, USA):
		\item November 20-21, 2014 (Paris, France:)
		\item September 17-18, 2015 (Washington, DC)
	\end{itemize}
	
	This collaboration also included a series of FIRE-GENI Research Experiment (FGRE) events, which
	provided training and interaction opportunities for student researchers and incubated a number of
	joint research efforts. FGRE events have been held in Ghent, Belgium on the following dates.
	\begin{itemize}
		\item July 7-11, 2014
		\item July 6-10, 2015
		\item July 11-15, 2016
	\end{itemize}
	
	Over twenty individual EU-US research collaborations have grown out of these collaborative efforts.
	In addition, a sizable body of open source software and emerging standards, including the Open Multinet
	federation software and ontology,\footnote{See \url{https://github.com/open-multinet}.}
	have arisen from this collaboration.
	\end{item}
	
	\begin{item}
	\textbf{SwitchOn Collaboration}
	The SwitchOn project explores collaborative opportunities in Future Internet research between the US and Brazil,
	with specific objectives of:
	\begin{itemize}
		\item Creating a mechanism to stimulate participation of US and Brazilian researchers in Future Internet research
		\item Providing coordination for high-impact research collaborative activities between the US and Brazil
		\item Exploring and identifying common interests in research and development to prepare for future large-scale collaborative research activities in Future Internet between the two countries
	\end{itemize}
	
	The SwitchOn collaboration organized workshops connecting US and Brazilian researchers,
	to further establish a global GENI presence capable of connecting researchers, end-users, 
	and all interested stakeholders at international scale through a fully federated infrastructure, as
	well as creating opportunity for  Brazilian researchers interested in collaborating in research into 
	global SDN networks, a topic of significant joint interest.
	\begin{itemize}
		\item January 8-9, 2015 (Miami, FL, USA)
		\item October 15-16, 2015 (S\~ao Paulo, Brazil)
	\end{itemize}
	
	\end{item}
	
	\begin{item}
	\textbf{Future Internet testbeds/experimentation between BRazil and Europe - EU (FIBRE-EU)}
	
	Beginning in 2011 with the FIBRE-EU project, and continuing today through the 
	Brazilian FIBRE testbed\footnote{See \url{http://fibre.org.br/}.},
	Brazilian and EU researchers have worked towards design, implementation
	and validation of a shared Future Internet research facility between Brazil and Europe, supporting the
	joint Future Internet experimentation of European and Brazilian researchers through four main activities:
	\begin{itemize}
		\item The development and operation of a new experimental facility in Brazil.
		\item The development and operation of a Future Internet facility in Europe based on enhancements and the federation (interoperability) of two existing FIRE infrastructures: OFELIA and OneLab.
		\item The federation of the Brazilian and European experimental facilities, both at the physical connectivity and control framework level, to support the provisioning of slices using resources from both testbeds.
		\item The design and implementation of pilot applications of public utility that showcase the power of a shared Europe-Brazil Future Internet experimental facility.
	\end{itemize}
	\end{item}
	
	\begin{item}
	\textbf{Japan-US Network Opportunity (JUNO) and JUNO2}	
	
	A series of Japan-US workshops on future networks that brought together researchers from both countries 
	from 2008 \cite{USJW2009}  to 2012 \cite{USJW2012} was the impetus for formally creating the first 
	US-Japan collaborative research program, JUNO, supported by the US National Science Foundation (NSF) 
	and the Japanese National Institute of Information and Communications Technology (NICT).
	
	The original JUNO focused on three topic areas:
	\begin{itemize}
		\item Network Design and Modeling
		\item Mobility
		\item Optical Networking
	\end{itemize}

	The second, program, JUNO2 is now underway. Several participants in GEFI 2018 are also
	engaged in JUNO2 and attended a co-located program meeting.
	The primary technical interest areas of JUNO2 include:
	\begin{itemize}
		\item Trustworthy IoT/CPS Networking
		\item Trustworthy Optical Communications and Networking 
	\end{itemize}
	\end{item}

\end{itemize}

\subsection{Opening Session}

Akihiro Nakao welcomed participants on behalf of The University of Tokyo, the host institution.
Prof. Nakao also served as general co-chair of the co-located IEEE CloudNet conference,
which took place during the three days immediately preceding GEFI 2018. He discussed the
relationship between the two events and potential collaboration opportunities.

Hiroaki Harai represented Japan's National Institute of Information and Communications (NICT).
He reminded participants of NICT's goals for GEFI and discussed the role of NICT
and the research testbeds supported by NICT. These testbeds include JOSE, StarBED,
RISE, and JGN. Dr. Harai looks forward to the integration of NICT's testbeds on JGN infrastructure.

Deep Medhi spoke on behalf of the US National Science Foundation (NSF).
He discussed a history of collaborations between NSF and NICT, as well as several
bilateral international collaborations between the US and other GEFI participant nations and regions.
These include JUNO2 (with Japan), cyberinfrastructure workshop collaborations (with Brazil), and
Internet Core \& Edge Technologies (ICE-T, with the EU). Dr. Medhi also highlighted a recent NSF
Dear Colleague Letter on mid-scale infrastructure, expressing a hope for interesting projects from
the US participants that address national- and international-level goals.

Mark Berman reviewed the workshop's goals to foster new international collaborations
for research infrastructure and testbed-supported research. He introduced the workshop structure,
reviewed the technical topic areas, and facilitated introductions among the workshop participants.

\subsection{Additional Information}

Participation in GEFI 2018 was invited via an open call for participation (CFP), circulated via relevant community
mailing lists.\footnote{CFP is available at \url{http://cloudnet2018.ieee-cloudnet.org/files/GEFI2018_CFP.pdf}.}
The CFP invited both session proposals and individual participant position statements. The GEFI organizing
committee identified session topics and participants with the goal of addressing important trends in research
where international collaboration can significantly increase scope, pace, and impact.

The GEFI 2018 workshop included five technical sessions, each of which is summarized in its own section
below. Supplementary information is included in section \ref{sec:detailed-information}.

Additional detailed information on the workshop, including speakers' presentation materials for the technical sessions,
is available from the workshop web page.\footnote{GEFI 2018 web page is: \url{http://indico.rnp.br/conferenceDisplay.py?confId=260}.}
A list of presentations, along with links to the presentation material, is found in section \ref{sec:presentation-summary}.

\section{Softwarization and virtualization of radios and networks}
Session summary was prepared by Abhimanyu Gosain (Northeastern University, USA) and Ingrid Moerman (Ghent University, Belgium).

\subsection{Session purpose}

\paragraph{Background}
 Interesting evolutions are happening at different layers that enable the creation of parallel isolated network slices, each slice forming a different independent network sharing the underlying wireless infrastructure and spectrum:
 \begin{itemize}
	\item At the networking level, Software-Defined Networking (SDN) decouples the network control and data plane forwarding functions, enabling directly programmable network control giving diverse network services to a variety of applications. Such approach allows a single physical network infrastructure to be virtualized into multiple and heterogeneous logical network domains, each domain serving a certain category of traffic flows in the most appropriate way. Such an approach is very encouraging, but has been originally designed for the wired domain and mainly involves the higher layers of the protocol stack (layer 4-7). It is primarily providing transport capacity and service differentiation up to the edge router. There is some recent work on SDN for wireless links (e.g. SDN-R project set up by ONF and in several H2020 5G projects), but these activities only consider current commercial wireless standards and do not exploit the full potential offered by more flexible radio platforms like SDR and future wireless technologies.

	\item At the radio level, we have observed the emergence of Software-Defined Radio (SDR). An SDR is a radio communication system where transceiver components that are typically implemented on hardware are instead implemented by means of software on a host computer or an embedded system equipped with programmable hardware like application specific instruction processor (ASIP) or field programmable gate array (FPGA).\footnote{Examples may include digital mixers, filters, equalizers, modulators/demodulators, multiple antenna techniques, etc. implemented in an application-specific integrated circuit (ASIC).} The concept of SDR is very encouraging for the development of state-of-the-art physical layer (PHY) functionality, because software programming allows much faster development cycles. The main problem with software environments is the slower sequential execution of algorithms, even when multi-core or many-core CPU (central programming unit) platforms or GPUs (graphics processing units) are used, in contrast to a very fast execution and a very high degree of parallelization in an ASIC, ASIP or FPGA. For this reason SDR development has so far mostly been limited to non real-time physical layer development or for latency-insensitive wireless communication, as software implementations do not always offer the fast execution times that are required for true networking experimentation (for example requiring fast acknowledgment of MAC frames within a few microseconds). Although SDR has gained a lot of interest because of the ease of software coding, we recently also observe an opposite trend to code more and more transceiver functionality on programmable hardware to achieve faster execution times.
\end{itemize}

As SDR and SDN research and developments are basically parallel evolutions happening in isolated research communities, interdisciplinary research efforts need to be stimulated to bridge the gap between SDR and SDN to realize true end-to-end networking with tailored QoS guarantees.

\paragraph{Session purpose and goals}
The purpose of this session is to explore the convergence of virtualization technologies for an end-to-end networked system in radio, transport, edge and cloud.
To solve the many research questions related to SDR-SDN integration, softwarization and virtualization, the session aims to:
\begin{enumerate}
	\item explore mechanisms for interdisciplinary collaborations and joint experimentally-driven research approaches
	\item set up interdisciplinary research teams that work closely together to discuss and define abstractions and (platform-independent) unified interfaces that maximally exploit the reconfigurable and programmable capabilities of radios and networks in a true end-to-end vision.
	\item provide access to open experimental facilities involving heterogeneous wired and wireless technologies that allow to set up realistic scenarios with sufficient level of scale (e.g. challenging dense wireless scenarios that have to share limited wireless spectrum in an efficient way while supporting different end-to-end services with diverging QoS requirements and security/privacy concerns)
	\item provide access to open softwarization and virtualization software platforms that can be easily extended with new functions, new abstractions and new/enhanced interfaces.
\end{enumerate}

\paragraph{Research problems}
The session explored the following questions:
\begin{enumerate}

	\item How to softwarize physical resources into real resources: how to represent physical radio/network resources (available in the physical domain) in the abstract domain or logical domain, and how to control and manage them by software at runtime (through a programmable framework). A real resource is an abstract representation of the physical resource, only considering general characteristics, rather than concrete realities. The abstraction manages the way in which systems interact, and the complexity of the interaction, by hiding details that are not relevant to the interaction. There may be different levels of abstractions: higher abstraction levels are more easy to use, but come at the cost of reduced flexibility and customization. 

Different abstraction levels further offer different levels of control, where we distinguish between parametric control and full composition of radio/network stacks by connecting/replacing/adding/removing functions.

	\item How to virtualize real resources: how to partition or aggregate real resources into virtual resources in order to create isolated end-to-end slices, each slice supporting a specific service and tailoring the real resources to a specific context of the service? How to realize joint wired and wireless network virtualization, exploiting full capabilities of flexible networks and radios (beyond commercial wireless standards of today)

	\item Some initial steps for radio softwarization and virtualization are taken at the European level in the H2020 projects WiSHFUL and ORCA that extend the radio data (or user) plane with a control plane offering runtime control through unified programming interfaces offering different level of abstractions on top of heterogeneous wireless technologies and hardware platforms. There are many more parallel activities worldwide on network softwarization and virtualization (SDN/SDX,NFV, etc.). But there is a lack of generalized models that would allow experimenters to fully exploit the capabilities of softwarization and virtualization of radios and networks without the need for deep knowledge of specific radio, network and hardware platforms.
	
	\item How to realize true end-to-end networking involving multiple network operator domains, each of them controlling one or more wired and/or wireless network segments? How different orchestrators, each of them controlling a separate network segment in an end-to-end connection will interact (east-west versus north-south interfaces)? Is there a need of macro-orchestrators capable to orchestrate the orchestrators of the separate segments? Which type of control strategies are needed: centralized, hierarchical, distributed? How to realize true end-to-end isolation of slices that involve multiple wired and wireless segments? How to monitor end-to-end performance in isolated slices? How to guarantee that a private network is really isolated from other networks (slices) sharing the same infrastructure? How to ensure end-to-end security and privacy in softwarized and virtualized networks? How to protect softwarized and virtualized networks against intentional or unintentional attacks (DDoS) and malicious applications/devices?
	
	\item How can research on softwarization \& virtualization be supported by experiments in realistic end-to-end scenarios? How to set up an end-to-end environment with sufficiently scale and heterogeneous network segments (wired, wireless, optical)? Or from another experimentation perspective, how can experimentation benefit from softwarization \& virtualization approaches. Note that virtualization and instantiation of independent slices sharing the same experimental infrastructure is already done for years (e.g. emulab, planetlab). Do we need/want to reinvent the wheel or can we better stimulate cross-fertilization between experimentation community and SDN/NFV research community?
\end{enumerate}

\subsection{Presentations}
This set of presentations includes four Europe, two US, two Japan and one Brazil presenters covering topics from advanced wireless radio testbeds, SDN integration, 5G Network slicing, convergent global scale end-to-end network system platforms as well as application vertical testbeds for Ultra reliable low latency applications.
\begin{itemize}
\item
\presentation
{Ingrid Moerman (imec -- UGent, Belgium)}
{ORCA vision on softwarization, virtualisation and end-to-end slicing}
{http://indico.rnp.br/getFile.py/access?sessionId=5&resId=6&materialId=0&confId=260}
{The H2020 ORCA project offers mature, real-time and versatile SDR platforms in advanced FED4FIRE compliant wireless test facilities. In this presentation the ORCA vision on end-to-end network slicing is presented for networks comprised of multiple network segments (wireless, wired, optical). It is explained that the creation of E2E network slices to provide guaranteed performance requires the slicing of each individual network segment, and the subsequent combination of these network segment slices. A hierarchical orchestration scheme, using a hyperstrator on top of the network segment orchestrators is proposed to deploy E2E network services.}

\item
\presentation
{Marco Ruffini (Trinity College Dublin, Ireland)}
{CONNECT's view on virtualisation: Testbeds, experimentation and future plans}
{http://indico.rnp.br/getFile.py/access?sessionId=5&resId=0&materialId=0&confId=260}
{The talk provides an overview of the testbed facilities and experimental activities on mobile and optical access network virtualisation carried out in the CONNECT research centre lab, in Trinity College Dublin. After reporting on individual experiments in the wireless and optical domains a use case is presented on the convergence of the fixed and mobile access network. Finally, a perspective is given on the wider involvement of CONNECT into the development of Dublin as a smart city and a glance at future plans.}

\item
\presentation
{Jerry Sobieski (NORDUnet A/S, Denmark)}
{The GEANT Testbeds Service: A ``Generic Virtualization Model'' and working code in a pan-European facility for network research}
{http://indico.rnp.br/getFile.py/access?sessionId=5&resId=2&materialId=0&confId=260}
{The GEANT Testbeds Service is a network research facility integrated with and co-located with the GEANT core footprint.  GTS implements a ``generic virtualization model'' that offers compute, transport, [SDN] switching, and storage in fully virtualized SDX network environments.  GVM/GTS is extending to support virtualized mobile edge resources. This lightning talk provides a quick view of the available facilities, collaborations, and a short list of interesting topics that could benefit from a global research approach.}

\item
\presentation
{Johann Marquez-Barja (imec -- UAntwerpen, Belgium)}
{Softwarization and Virtualization as a mean for convergence and interoperability}
{http://indico.rnp.br/getFile.py/access?sessionId=5&resId=5&materialId=0&confId=260}
{It is fact that the use of software toolkits deployed on top of the communication network components enable flexible functionalities capable to dynamically reconfigure the network. Such reconfigurability opens up a new dimension of possibilities to deliver connectivity. However, in order to deliver end-to-end connectivity several challenges need to be addressed. The H2020 EU-Brazil FUTEBOL project has developed and deployed a set of software toolkits, wrapped-up in a Control Framework, that is capable to converge Wireless, Optical, MEC, and IoT network components in order to truly provide end-to-end connectivity. This talk addresses some challenges related to the convergence and interoperability of heterogeneous network segments; the FUTEBOL control framework; and the ongoing experimentation between Brazil and Europe Testbed facilities.}

\item
\presentation
{Ivan Seskar (Rutgers University, USA)}
{Integrating Implementation Frameworks for Edge Network Applications}
{http://indico.rnp.br/getFile.py/access?sessionId=5&resId=1&materialId=0&confId=260}
{This talk focuses on the challenges of integrating and harmonizing virtualization across multiple implementation technologies (CPU, GPU and FPGA). We currently lack a framework to optimize the placement of latency critical application workloads. Considering an example of a real-time AR application with low-latency constraints running on a next-generation wireless edge cloud network, we can identify several opportunities for cross-layer optimizations that are necessary for realizing extremely low ms level application latencies.  We pose the question of whether it is possible to continue independent development of SDR, SDN, NFV and cloud, or whether there is a better way to integrate and harmonize for end-to-end orchestration and QoS support while the standards are still at an early stage.}

\item
\presentation
{Abhimanyu Gosain (Northeastern University, USA)}
{Platforms for Advanced Wireless Research}
{http://indico.rnp.br/getFile.py/access?sessionId=5&resId=4&materialId=0&confId=260}
{This talk introduces PAWR, \$100M US public-private research partnership to support creation of four city-scale experimental platforms for advancing fundamental wireless research. This program plans to blend cutting-edge wireless and cloud innovations with a large scale geographical community. We share our vision of creating a network platform architecture, which is extremely agile, dynamic, cost-effective, adaptable, sliceable and extensible, giving unprecedented programmability and control to a broad set of researchers. The talk also details the first two Platforms addressing mmWave and Massive MIMO radio technologies and their requirements from a cloud infrastructure perspective.}

\item
\presentation
{Jos\'e Rezende (RNP, Brazil)}
{SDI and Elastic Optical Testbed at RNP}
{http://indico.rnp.br/getFile.py/access?sessionId=5&resId=7&materialId=0&confId=260}
{In this talk, we report on the SDI project that is being conducted at RNP/Brazil with the aim of offering flexible and value-added services for its customers, incorporating as much automation as possible. The Software-Defined Infrastructure consists of an overlay network and a distributed edge cloud. In addition, plans are presented to deploy an elastic optical testbed for remote experimentation. }

\item
\presentation
{Aki Nakao (University of Tokyo, Japan)}
{Application Specific RAN Slicing}
{http://indico.rnp.br/materialDisplay.py?sessionId=5&materialId=0&confId=260}
{We introduce our recent research on in-network data analytics and deep machine learning in softwarized infrastructure, especially for emerging new generation mobile networking. Advanced traffic data analytics becomes possible because not only fixed transport equipment but also mobile base stations are getting softwarized, facilitating the deployment of complex data processing within network by means of data-plane programmability. In fifth generation mobile networking (5G), enhanced Mobile Broadband (eMBB) , Ultra Reliable and Low Latency Communication (URLLC) and massive Machine Type Communication (mMTC) are expected to be serviced without interference. While eMBB is planned to be in service soon, we posit that URLLC services such as cooperative driving, drone surveillance, etc. become the next focus in near future, and require ``mobile network slicing via advanced traffic classification'' and ``extreme edge computing.'' We introduce in-network data analytics and deep machine learning in mobile networking to enable per-application end-to-end fixed mobile network slicing for upcoming 5G infrastructure.}

\item
\presentation
{Toshiyuki Miyachi (NICT, Japan)}
{Wireless Emulation on StarBED}
{http://indico.rnp.br/getFile.py/access?sessionId=5&resId=3&materialId=0&confId=260}
{NICT has developed and operated StarBED, a large-scale general purpose network testbed since 2002. In this talk, NICT presents how the function of wireless emulation is added to its wired network.
\end{itemize}

\subsection{Discussion Details}

\subsubsection{Ambitions and goals}

The principal ambition of the session was to explore potential strategies for most effectively leveraging open-source software and hardware platforms in future large-scale wireless networking research and research infrastructure development activities. Other goals included:
\begin{enumerate}}

	\item Establish and document a baseline catalog of current open-source project work in relevant technical areas, including wireless networking software and hardware, cloud computing, Software Defined Networking (SDN), Network Function Virtualization (NFV) and experimental research management and support infrastructure. 
	\item Conduct a gap analysis for the aforementioned work across parameters such as: availability of code, community engagement, and reliance on adopted standards, features, tools and accessibility. 
	\item Evaluate the feasibility of developing a single end-to-end, open-source reference model that ties together the three technical areas to support near-term standardization efforts as well accommodating blue-sky research on large-scale research platforms. Main research challenges related to softwarization and virtualization of radios and networks to realize a true end-to-end vision.\footnote{See also white paper on ORCA vision: \url{https://orca-project.eu/wp-content/uploads/sites/4/2018/10/orchestrating_e2e_network_slices_Final.pdf}}
	\item Potential joint/interdisciplinary experiments and showcases that can be shared with and reproduced by the research community: scenario for experiment, scale of experiment, key expertise required. What can be offered to the research community: software tools, data sets, best practices, how to share?
\end{enumerate}

The following ``big ideas'' were generated:
\begin{enumerate}
	\item How do we develop mechanisms to incentivize the intersection of the radio and networking community? User expertise (also determines the size of the community) can be divided into Basic and Advanced. Lowering the entry point increases the number of users (e.g. you do not need VHDL knowledge to use FPGA, you can offer interfaces to orchestrate radio functions on FPGA).
	\item Creation of joint end-to-end cross-segment (wired, optical, wireless) experiments. Different segments have different compute performances (ASIC to CPU/GPU in radio), latencies (ms for Ethernet versus micros for optical) and dynamics (fixed capacity in wired networks versus very variable capacity in wireless networks). These experiments are enabled by definition and standardization of open interfaces (between hyperstrator and individual segment-specific orchestrators). The meta-output of such experiments is to understand the latencies (end-to-end is a composition of latencies in different segments), the need  novel monitoring approaches (not only at the segment level, but also in-network cross-segment monitoring techniques) and also gather requirements for developing such abstractions from top to bottom, and from bottom to top. This comes back to bringing stove piped communities together.
	\item One big area in which open-source software for mobile networking research could be improved is the creation of a software suite that takes a more modular, library-like approach to wireless network creation. While many existing software stacks are ``modular'' in the sense that they consist of several 3GPP components, and it is possible to modify or replace individual components, this is a very specific form of modularity that makes strong assumptions about the needs of the network. This limits research innovation to specific well-known ``shapes,'' and brings with it a large learning curve, since an experimenter must understand a large number of 3GPP standards to begin building the simplest network. A toolkit that is decomposed into much smaller pieces - essentially, a set of technical building blocks - would give experimenters a much easier time thinking outside of the 3GPP ``box'', and would facilitate the creation of much simpler ``minimum working networks.''  A clean separation of the technical building blocks from the interfaces that implement standards is paramount. A technical building block may be used across multiple standards, multiple generations of the same standards. Standards are largely about "interfaces". This idea is one step further from the current ``modular, library-like approach.'' A good analogy here is the Click modular router: the simplest working router one can build with Click simply forwards all traffic from its input ports to its output ports. From there, the user can build up more complicated configurations, using standard IP tools such route lookup, TTL check-and-decrement, checksums, etc., and can easily write their own modules to replace or complement the standard ones. Building this kind of architecture for the mobile network space would be a significant challenge, but one that is likely to pay off in terms of more transformative research.
\end{enumerate}

Challenges include the following:
\begin{enumerate}
	\item Apart from the obvious technical challenges, the relationship between sharing artifacts and software design methodologies is a major issue. In many ways, the intention of sharing software and encouraging its modification (both critical in the research community) impacts the way implementation is done. In light of this, it is necessary to tackle issues about modularity to appropriately support new APIs, and to encourage research that departs significantly from todayÕs networks.  
	\item The idea of modularity is always associated with well-documented and well-maintained artifacts. So, the challenge in this direction is not only how to improve the modularity of open-source software for wireless networks, but how sure that the projects have the discipline to keep high-quality, up-to-date documentation.
	\item Interestingly, the idea of network softwarization enables network programmability, flexibility and extensibility. In order to provide interoperability among heterogeneous networks, platforms and experiments as a set of extensible APIs is necessary. The main point is how to specify and design such APIs and in the same time support hardware software co-design. From this point of view, Industry and Academia have a fundamental role.  
	\item Regarding the management and orchestration of network resources via software the existing ones already includes several requirements. The issue here is how to integrate these existing tools to provide the needed level of management and orchestration. Taking into account the network testbeds context, how can we provide flexibility at the experimental layer through the existing tools? The start point could be through some degree of standardization.
\end{enumerate}

The following next steps were identified:
\begin{enumerate}
	\item Survey the testbed landscape for features, scale, experimentation orchestration framework(s) and size of user base across all technical areas (radio, transport, edge and cloud). The group will develop, execute, analyze and publish the results with the broader community. 
	\item Continue participation in industry sponsored ``open'' activity forums such as TIP, O-RAN and OpenAirInterface and disseminate information about large/mid-scale research infrastructure activities in the academic domain.
\end{enumerate}

\subsubsection{Consensus and Open Issues}
A clear common theme that emerged from the session presentations included;
\begin{enumerate}
	\item A clear definition, separation, distinction around terminology for softwarization, virtualization, and NFV (MEC/Cloud) and its utility in the mobile networking landscape.
	\item Need for hypervisor = super-orchestrator of top of segment orchestrators for MANO in end-to-end networked systems.
	\item Network slicing techniques from the perspective of RAN still requires more attention from both academia and industry.
	\item Leverage fine-grained expertise in different research communities and utilize for well-defined and open interface between the hyperstrator and the underlying orchestrators.
	\item MIMO and mmWave relevance for the medium term. These technical areas also cross pollinate requirements and new challenges in the northbound technical areas For e.g. availability of high speed, low latency optical interconnection, offers a lot potential for huge processing needs involved in MIMO and mmWave.
\end{enumerate}
	
\subsubsection{Potential collaborations}
The projects and research platforms listed below demonstrate the efforts of the GEFI community as well as the diverse set of solutions deployed: In Europe -- ORCA, 5GinFire, 5Tonic, 5GIC at Univ. of Surrey, OneLab, FIT, Fraunhofer 4/5G Testbeds, Bristol is Open, OpenAirInterface, Fed4FIRE, FIWARE Lab Node, 5G-EmPOWER at CREATE-NET, and imec iLab.t testbeds. In the US -- PAWR, Wiser-Lab, WiTEST-Lab, DETER, Wireless@VT, WINGSNet, WARP, PhantomNet and ORBIT. In Japan -- JOSE and StarBED and RNP in Brazil. In EU and Brazil -- FUTEBOL.

One area for collaboration is the design, specification and eventually operation of a multi-platform experimentation infrastructure that is itself modular in the sense that it can be constructed as building blocks. This will enable academic or research groups to start from the desired granularity of the testbed environment and scale up as needed. 

Such an approach implies the need for a blueprint (meta) architecture that facilitates integration and interoperation of various network components and various network segments across the end-to-end path. The challenge here is the need to address an unavoidable trade-off between standardized architectures and research flexibility as a precondition for innovation and new ideas. A possible starting point is the ETSI NFV reference architecture as it ensures convergence with key standardization activities. From the perspective of an experimentation infrastructure the blueprint architecture needs to define interfaces that could contain a subset of existing architecture standards of major sub-components of the end-to-end path. The interfaces and API specifications are crucial for experimentation deployments across the infrastructure. 

Another area is the specification or extensions of domain specific languages (e.g., P4) or modeling languages (such as YANG and TOSCA) that enable network service descriptions that could drive automated experimentation deployment. Combined with a portal/repository for VNFs and testbed setup and configuration tailored to specific research needs, this orchestration will contribute towards an organized platform ecosystem for researchers.

\section{Networking Experiments}
Session summary prepared by Kate Keahey (University of Chicago, USA) and Lucas Nussbaum (Université de Lorraine, France).

\subsection{Session purpose}

Networking is obviously a central concern when performing experiments in the distributed systems context.
Specifically, experiments range from network-focused experiments, such as designing and evaluating protocols, to higher-layer experiments, requiring the interconnection of large number of resources by a controlled or isolated network.
Various services have been designed and provided by testbeds over the years covering:
\begin{enumerate}
	\item on-demand isolation and interconnection of resources, both inside testbeds, and between testbeds (with federated approaches such as SDX);
	\item network emulation services, to provide specific network conditions (limited bandwidth, added latency, faults, etc.).
 \end{enumerate}
However, new use cases are emerging, causing a shift of requirements. One could mention edge computing scenarios that require combining various kinds of geographically distributed devices in a controlled environment, something that is not traditionally available. Another example is HPC / Big Data / AI scenarios, that require providing the same level of control for HPC network technologies (InfiniBand, Omni-Path).

The goal of this session is to bring together experimenters and testbed operators to understand and explore: (a) use cases for advanced networking experiments, and their requirements; (b) the current state of support for networking experiments, both at the testbed level, and at the testbed federation level.

\subsection{Presentations}
\begin{itemize}

\item 
\presentation
{Violet R. Syrotiuk (Arizona State University, USA)}
{Experiments in Wireless Networking}
{http://indico.rnp.br/getFile.py/access?sessionId=6&resId=11&materialId=0&confId=260}
{This talk reported on experiments in wireless networking performed on the
\textsl{w-ilab.t} testbed. Several interesting issues were encountered and
overcome, both on the level of the experiment setup and orchestration, and
of the experiment design.}

\item
\presentation
{Brecht Vermeulen (Ghent University -- imec, Belgium)}
{Flexible testbed infrastructure: from pure networking experimentation towards general experimentation}
{http://indico.rnp.br/getFile.py/access?sessionId=6&resId=1&materialId=0&confId=260}
{Using flexible testbed infrastructure and tools to evolve from pure
networking experiments for expert researchers 15 years ago to a mix of
networking, (GPU) computing, storage, IoT and wireless experimenting for
starting and expert researchers and student classes today.}

\item
\presentation
{Paul Ruth (RENCI, UNC -- Chapel Hill, USA)}
{Software Defined Networking and Wide-Area Stitching on NSF Cloud Chameleon}
{http://indico.rnp.br/getFile.py/access?sessionId=6&resId=4&materialId=0&confId=260}
{The NSF Cloud Chameleon is nearing the end of the first year of its second
phase.  A key thrust of this phase is to enable advanced networking
experiments. Toward this goal, members of the ExoGENI team at RENCI were
added to the project.  This presentation shares some recently released
networking features and the experiments that they enable. The new features
include software defined networking, 100 Gbps wide-area experiments, and L2
stitching to other testbeds. Central to this work is the deployment of new
Corsa DP2000 series switches which allow deeper programmability of the
network enabling us to provide isolated OpenFlow networks controlled by the
user. These capabilities are currently deployed in the production Chameleon
testbed and we will share both our experience deploying these features as
well as the initial user feedback on useability. These capabilities both
enable advanced networking experiments on Chameleon as well as enabling
experiments that span Chameleon and other testbeds.}

\item 
\presentation
{Paul Ruth (RENCI, UNC -- Chapel Hill, USA)}
{Toward Inter-Testbed Experimentation: Programmable Core Networks}
{http://indico.rnp.br/getFile.py/access?sessionId=6&resId=5&materialId=0&confId=260}
{Cloud and network testbeds have a wide range of capabilities, strengths, and
		weaknesses. Each testbed not only contains different types of hardware but
		also has different ability to scale, different geographic location(s), and
		domain affiliation(s).  Some testbeds are large centralized clouds, others
		are small edge clouds. Some are shared, others are private. Relying on any
		single testbed limits the experiments that can be performed. Ideally users
		could combine heterogeneous testbeds from multiple domains, however
		currently this is difficult at best and often impossible.  In order to
		support a wider range of experiments it is important that the future
		network testbeds provide mechanisms for users to combine resources from
		various sources. Efforts, like the GENI federation in the U.S. and others
		internationally, have progressed toward creating common APIs implemented by
		multiple edge cloud testbeds. However, we still rely on ad hoc methods for
		connecting and deploying experiments across un-federated testbeds. Further,
		physical connectivity between testbed sites typically relies on the shallow
		programmability of stitched point-to-point circuits directly from one edge
		to another.   These limitations are a result of most deeply programmable
		cloud and networking testbeds being designed as edge clouds that lack a
		deeply programmable core network. In response, a group of U.S. researchers
		have proposed that the NSF develop an experimental infrastructure composed
		of a Future Core Network (FCN) and a Future Edge Cloud (FEC).  The proposed
		FCN/FEC network will form a deeply programmable substrate interconnecting a
		variety of national and international research infrastructures as well as
		public cloud providers. It will sit between current cyberinfrastructure
		investments in campuses, public clouds, NSF Clouds, PAWR Wireless Edge,
		shared research networks, and shared computing facilities. The key
		contribution is that the both the edge and the core are programmable and
		connectivity to the core is standardized enabling researchers to include
		private instruments and other resources. Any programmable core network
		testbed should be designed with international collaboration in mind. The
		FCN proposed in the U.S. comes from the collective experience and insight
		of many researchers involved in GENI.   This presentation will share the
		topics and ideas from this paper with the GEFI workshop as well as gain
		additional insight from potential international collaborators.}

\item
\presentation
{Anirban Mandal (RENCI, UNC -- Chapel Hill, USA)}
{Investigating Use of Distributed Research Infrastructures and Testbeds for Domain Science Experimentation and Validation}
{http://indico.rnp.br/getFile.py/access?sessionId=6&resId=8&materialId=0&confId=260}
{Recent advances in dynamic, networked cloud infrastructures like NSF GENI and
Software Defined Networking (SDN), provide the building blocks to construct
such integrated, reconfigurable, end-to-end infrastructure that has the
potential to increase scientific productivity and support domain science
experimentation. In addition to supporting computation, storage and data
movements for large-scale domain science collaborations, distributed research
infrastructures and testbeds are rapidly becoming virtual labs for
experimenting with novel algorithms, models, and data management approaches for
different domain science applications. Hence, we envision that global,
distributed testbeds will become essential building blocks and testing grounds
for the development of research infrastructures for next generation domain
science experimentation.

In our prior work, we have integrated scientific workflow systems like Pegasus
with dynamic resource provisioning on ExoGENI, which made it easier for science
workflows to leverage dynamic infrastructures. We have also experimented with
the suitability of executing domain science applications on the GENI testbeds.
We plan to continue our efforts in enabling the use of advanced, distributed,
research infrastructure and testbeds for domain science experimentations with
focus on observational sciences in the context of a recent award. Another novel
use of distributed, testbed infrastructures is for evaluation and validation of
distributed science applications and workflows as demonstrated in our recent
work on the Panorama 360 project, where we performed network performance
analysis of workflow data transfers for scientific applications by actuating
systematic network perturbations on ExoGENI testbed. Hence, designing
distributed research infrastructures that can provide an experimental platform
to validate domain science applications and help develop methods to analyze
faults is of utmost importance. We are also using research testbed
infrastructures for studying the integrity and reproducibility of domain
science applications when faced with different kinds of intentional or
unintentional attacks that threaten the validity of scientific results. As part
of a recent award, we are planning on using distributed testbeds and anomaly
injection software to train machine learning algorithms for automatic detection
of domain science integrity failures.

Hence, our position is that distributed global testbeds and research
infrastructures are essential for not only executing different kinds of domain
science applications at scale, but also for evaluating, validating and
experimenting with domain science applications.}

\item
\presentation
{Cesar Marcondes (Aeronautics Institute of Technology, Brazil)}
{Proposal of Experimental Environment for Data Plane in Computer Networks}
{http://indico.rnp.br/getFile.py/access?sessionId=6&resId=3&materialId=0&confId=260}
{Testbeds are fundamental platforms for the creation and validation of new
technologies and Internet architectures of the future. They enable a more precise,
scalable and close control of the real characteristics of a network like the Internet
environment. In this way, these testbeds allow the validation of architectures
elaborated from scratch with new concepts based on new technologies and focused
mainly to solve all the security and performance implications that clutter this
current model of the Internet. To fill this gap, this work aims to propose an
environment where the experiment has the possibility to control the data plan in
networks, using an API that reduces the time of creation and instantiation of the
experiment, without functional restrictions to the user, aiming to provide a flexible
environment for experimenting with significant parts operating on the Internet.}

\item
\presentation
{Cesar Marcondes (Aeronautics Institute of Technology, Brazil)}
{MiniSecBGP: Lightweight Security BGP Emulation Testbed}
{http://indico.rnp.br/getFile.py/access?sessionId=6&resId=2&materialId=0&confId=260}
{The Border Gateway Protocol (BGP) is the routing protocol responsible for connecting
the entire Internet and attacks on BGP systems have the potential to cause
significant financial impact or even jeopardize the sovereignty of a country. Despite
their relevance, it is not common environments that support in-depth studies of BGP.
In this line, this work presents the MiniSecBGP, a testbed that supports the
emulation of part of the internet topology in a realistic mode, by interacting with
widely adopted BGP implementations. MiniSecBGP has a modular architecture, making it
flexible and expandable. Preliminary tests indicated good scalability and accuracy of
the proposed solution.}

\item
\presentation
{Joe Mambretti (Northwestern University, USA)}
{Global Federated Network Research Testbeds: Trends and Opportunities for Collaboration}
{http://indico.rnp.br/getFile.py/access?sessionId=6&resId=6&materialId=0&confId=260}
{The International Center for Advanced Internet Research (iCAIR) at Northwestern
University is currently supporting over 20 major experiment networking testeds,
most are national and international in scale (and federated). I think it would
be useful to give a presentation on the future direction of such testbeds given
current and emerging research trends and requirements, especially as related to
programmable networking.}

\item
\presentation
{Jiang Liu (Beijing University of Posts and Telecommunications, China) -- talk presented by Jason Liu}
{China's Future Network Testbed}
{http://indico.rnp.br/getFile.py/access?sessionId=6&resId=7&materialId=0&confId=260}
{China is building an open, easy-to-use and sustainable future network testbed,
including basic network in-frastructure, operation control centers,
experimental platforms, supporting applications and innovative architectures.
The facilities will span 40 cities in China, over 88 backbone network nodes and
133 edge networks. The backbone link bandwidth will reach 100G, with
interconnection with domestic and international networks and other future
network testbeds. The testbed is projected to support no less than 128
heterogeneous networks and 4096 concurrent experiments. The main control center
will be built in Nanjing, and three regional centers will be built in Beijing,
Hefei and Shenzhen to achieve efficient operation and management of the test
facilities. The Nanjing control center is responsible for monitoring the
operational status of the entire network, maintaining a unified view of the
network, running the management services, and coordinating the network
resources to provide users with a unified service.}

\item
\presentation
{Srivatsan Ravi (University of Southern California, USA)}
{Cyber-experimentation for secure and distributed software-defined networking infrastructures}
{http://indico.rnp.br/getFile.py/access?sessionId=6&resId=9&materialId=0&confId=260}
{Software-Defined Networking (SDN) outsources the control over the data plane to
logically centralized software called the control plane, thus essentially
raising the level of abstraction for network programming. The control plane
allows for expressing and composing network policies of varying networking
applications and translating these to rules installed onto the switches for
handling network flows. A provably robust and secure SDN control plane can
provide a clean separation of the networking data plane, thus, naturally
enforcing protection of networked infrastructures against an ever-increasing
attack surface resulting from a rapidly evolving hardware-software ecosystem
and malicious networking applications. This design will potentially have a huge
impact on smart cities and smart transportation systems relying on dispersed
edgecomputing infrastructures incorporating heterogeneous edge devices.
However, protocols for consistent network policy updates in distributed SDN
deployments are heavily dependent on the dimensionality of the distributed
computation space specified by the adversarial model, application semantics,
etc. This position paper motivates testbed infrastructure development for
cyber-experimentation with security policies and mechanisms enabling resilient,
secure, cross-domain SDN networking and services involving multiple domains of
authority.}

\item
\presentation
{Dongkyun Kim (KISTI, Korea)}
{SDN-based Orchestration of Wide-Area Virtual Networks and Distributed Resources on KREONET}
{http://indico.rnp.br/getFile.py/access?sessionId=6&resId=14&materialId=0&confId=260}
{This talk will present SDN-oriented computing, storage, networking
orchestration architecture based on virtualized container networking. The
architecture integrates distributed service resources using Kubernetes and
wide-area virtual networking resources using VDN (Virtually Dedicated
Network) application on KREONET-S which is an OpenFlow native SDN-WAN for
R\&E community in South Korea. The orchestration system allows KREONET-S
users to dynamically and rapidly manage their demanding containerized
computing and storage resources coupled with high-performance virtually
dedicated networks activated for high speed, low or zero packet loss and
optimum end-to-end (or edge-to-edge) latency.}

\item
\presentation
{Leandro Ciuffo (RNP, Brazil)}
{FIBRE 2.0}
{http://indico.rnp.br/getFile.py/access?sessionId=6&resId=10&materialId=0&confId=260}
{This talk presented the FIBRE 2.0 plans, and a vision for a national platform
for experimentation on cloud/edge computing, IoT and SDN.}

\end{itemize}

\subsection{Discussion Details}

It is clear from the presentations and the discussions that there is now a wide
offering of testbeds for experiments related to networking, ranging from
wireless (e.g.; 5G testbeds) and IoT testbeds to cloud testbeds. They offer a
wide range of features, with various levels of configurability,
programmability, and isolation. There are also several efforts on providing
various kinds of interconnection between testbeds, using technologies such as
VLAN stitching and Software Defined eXchanges. There might be a need to \emph{clarify
those offerings and the surrounding categorizations and terminology; a survey}
written in collaboration by the GEFI community would be a very nice outcome.

There seem to exist two important gaps in the current offering. First, in order
to address new use cases in Fog/Edge computing, there is a need for \emph{testbeds
that bring together both wireless/IoT resources and Cloud resources} in a
coherent whole, or for testbeds on both sides to collaborate and build a
sufficiently close relationship to provide an integrated service.

Second, while configurability of networking resources and isolation at the
logical level (VLAN-like isolation) are widely available, there are very few
testbeds that \emph{provide performance guarantees at the networking level},
such as bandwidth guarantees, at the networking level to experimenters. This is
in contrast to the situation for nodes, where it is common for users to get a
dedicated hardware node. It will be interesting to explore which technologies
and tools could be leveraged to provide performance guarantees at the
networking level, without vastly over-provisioning the infrastructure. The same
issue also arises at the level of testbeds federations and wide-area testbeds
(and is even worse, due to the use of network links with smaller capacity, that
are usually shared by many experiments.

Two side issues were raised during the discussions. The first one is the need
for a \emph{better understanding of classical Design of Experiment methods}, and of
how they can be applied (and maybe adjusted) to our field, and then made widely
known to experimenters. This is probably especially true for networking
experiments due to the large number of factors that can affect an experiments'
results, and the cost of exploring all those factors systematically.

Another side issue, related to the large and costly infrastructures that are
required for such experiments, is the problem of \emph{sustainability}, and of
seeking funding models that enable building large infrastructures instead of
sets of smaller infrastructures.

\subsubsection{Potential collaborations}

\begin{itemize}

	\item There are plans, coordinated by Lucas Nussbaum, to work on a survey of
		\textsl{cloud} testbeds, involving at least Chameleon (Kate Keahey, U.
		Chicago), CloudLab (R. Ricci, U. Utah), Grid5000 and Starbed (T. Miyachi,
		NICT). Its scope could be revised depending on who is motivated
		to work on it.
		
	\item Stephen Schwab and Srivatsan Ravi identified collaboration on extending
		ongoing work on wide-area SDN experimentation in DETER with Jerry
		Sobieski of NORDUnet.

\end{itemize}

\section{EdgeNet}
Session summary prepared by
Rick McGeer (US Ignite, USA),
Timur Friedman (Sorbonne Universit\'e, France),
and 
Aki Nakao (University of Tokyo, Japan).

\subsection{Session purpose}

A key recommendation of GEFI 2017 was that a joint project between the EU, Japan, Brazil, and the US be undertaken to build a worldwide edge cloud as a joint project of the four participating entities and as a platform for future experimentation and collaboration.  The proposed participants undertook that project and have built a prototype of the system, EdgeNet, (https://www.edge-net.org) and propose to put this under the governance of a worldwide consortium, modeled on the highly-successful PlanetLab consortium.  EdgeNet is designed to avoid the pitfalls that have dogged earlier edge clouds: bespoke software, reliance on dedicated hardware and  special-purpose control frameworks.  EdgeNet is a virtual infrastructure, designed to spread easily and live lightly on the land; all a site needs to dedicate is a virtual machine with a routable IPv4 or IPv6 address.  EdgeNet uses industry-standard software wherever possible, leveraging both industry-wide maintenance and the plethora of available training materials.  

This session, reviewed the current status of EdgeNet, expansion plans in each of the four GEFI regions.  Scientific questions addressed included establishment and maintenance of a global edge cloud, applications of a global edge cloud, new science enabled by a global edge could, and new techniques to deal with churn of the underlying infrastructure.

\subsection{Presentations}

\begin{itemize}

\item
\presentation{Rick McGeer (US Ignite, USA)}
{EdgeNet Introduction}
{https://drive.google.com/open?id=1SiF5rW8rsPg-h2IOhd6xA38WIBZU3IUt}
{This talk describes EdgeNet, a lightweight cloud infrastructure for the  edge.
We aim to bring as much of the flexibility of open cloud computing as
possible to a very lightweight, easily-deployed, software-only  edge infrastructure.

EdgeNet has been informed by the advances of cloud computing
and the successes of previous  distributed edge clouds: PlanetLab, GENI, G-Lab, SAVI, and
V-Node.  Each of these had a large number of small points-of-presence,
designed for the deployment of highly distributed experiments and
applications.  EdgeNet differs from its predecessors in two significant
areas: first, it is a software-only infrastructure, where each
worker node is designed to run part- or full-time on existing
hardware at the local site; and, second, it uses modern, industry-standard
software both as the node agent and the control framework.  The
first innovation permits rapid and unlimited scaling: whereas GENI
and PlanetLab required the installation and maintenance  of dedicated
hardware at each site, EdgeNet requires only a software download,
and a node can be added to the EdgeNet infrastructure in 15 minutes.
The second offers performance, maintenance, and training benefits;
rather than maintaining bespoke kernels and control frameworks, and
developing training materials on using the latter, we are able to
ride the wave of open-source and industry development, and the
plethora of industry and community tutorial materials developed for
industry standard control frameworks.  The result is a global
Kubernetes cluster, where pods of Docker containers form
the service instances at each point of presence.

EdgeNet is a joint project of US Ignite, the  NYU Tandon School of Engineering,
PlanetLab Europe, UC-Berkeley, the University of Victoria (Canada),
and is open to additional collaborators. It currently features over 30 sites
across the US, Canada, and France, and is poised for rapid expansion.}

\item
\presentation{Ada Gavrilovska (Georgia Institute of Technology, USA)}
{Distributed Systems Class Projects with EdgeNet}
{https://drive.google.com/open?id=1XnaLXfk_ML85_o2bBZTvfINAGiNWcVWs}
{This talk described serveral use cases for an edge cloud:

\begin{enumerate}
\item Analytics for autonomous vehicles
\item Analytics for IoT. 
\end{enumerate}

This breaks down into a number of subproblems and issues.

Third parties would likely be involved in providing these analytics, which raises security issues.  AirBox is  a platform that provides containers built on Intel SGX, which provides a protected/secured mode for selected parts of an  application that requires container security.

Georgia tech is working on IoT management support.  The edge is close to where IoT objects are under attack.    Experiments show lower detection time and ability to remove significant amounts of attack traffic by deploying to the edge. This is one example  of an experiment that could be run in EdgeNet.

A second area of experimentation is a federated mode or analstics,  without centralized learning -- collaborative learning, using a large number of federated edge nodes.

A third example is   visual computing, specifically image and video processing. Models can be run  at the edge, but  there is a wide disparity in hardware available at the edge, so  some flexibilityis needed  in what sort of models to be run at the edge.

Each of these experiments are examples of IoT analytics experiments which would benefit from a ubiquitous edge cloud such as EdgeNet.
}

\item
\presentation{Yuuichi Teranishi (NICT, Japan)}
{Cross-layer Dynamic Data Flow Processing on Edge Clouds}
{https://drive.google.com/file/d/1Opvab8PhfY701fcB723H6OHJw6OKtI9K/view}
{A testbed operated by NICT in Japan called JOSE (Japan-wide Orchestrated Smart/Sensor Environment), which provides distributed edge cloud testbed facilities was introduced. There are 5 distributed data centers in Japan (Yokosuka, Kanazawa, Kyoto, Tokyo, Osaka) and more than 10,000 virtual machines (or containers) are available for the experiments on these data centers. The storage / computation / network resources are controlled and virtualized by SDN and SDI functions in a centralized manner. Each experimenter can deploy their processing modules on the distributed servers in the dedicated network slice. As one of the next steps of such testbed technologies, functions based on ``IoT Cross-Layer Edge Computing Architecture'', a two-layered (Platform as a Service layer and Infrastructure as a Service layer) dynamic edge cloud architecture, are currently being developed. A typical application scenario is a video object detection processing on an edge cloud, in which multiple edge resources are dynamically allocated to cope with the resource limitations and demands for the small processing latency.}

\item
\presentation{Srivatsan Ravi and Stephen Schwab (University of Southern California Information Sciences Institute, USA)}
{Research infrastructure and experimentation towards a smart, secure and scalable edge}
{https://drive.google.com/open?id=1aTWlWCyyVwSMTGUUNpr_HGArzzZrs7S7}
{This talk describes DCompTB which is a network testbed facility that allows researchers to rapidly design, deploy and execute complex experimental networked systems. Experiments involving hundreds of nodes can be materialized in minutes. The compute and network platforms provided by the testbed are open source hardware, providing maximum flexibility as an experimental platform. All nodes are have remotely accessible UART consoles for low-level systems development. The testbed provides advanced network modeling and emulation capabilities, allowing researchers to deploy systems in high operational-fidelity edge environments. This talk describes DCompTB which is a network testbed facility that allows researchers to rapidly design, deploy and execute complex experimental networked systems. Experiments involving hundreds of nodes can be materialized in minutes. The compute and network platforms provided by the testbed are open source hardware, providing maximum flexibility as an experimental platform. All nodes are have remotely accessible UART consoles for low-level systems development. The testbed provides advanced network modeling and emulation capabilities, allowing researchers to deploy systems in high operational-fidelity edge environments. DCompTB leverages the Merge software architecture~\cite{merge-paper} to dynamically integrate disparate testbeds in a logically centralized way that allows researchers to effectively discover, and use the resources and capabilities provided the by evolving ecosystem of distributed testbeds for the development of rigorous and high-fidelity cybersecurity experiments.}

\item
\presentation{Timur Friedman (Sorbonne Universit\'e, France)}
{EdgeNet and PlanetLab Europe}
{https://drive.google.com/open?id=1AIiqFrhqQN2kBcj3OtQuIkp97e2-kuDY}
{PlanetLab Europe remains an active infrastructure, with  343 nodes at 205 sites.Through federation with other testbeds, PlanetLab Europe users have access to varied resources on over a thousand nodes located at more than 500 sites worldwide.  However, the PlanetLab Europe software and hardware inrastructure present an   increasing maintenance challenge, due to the ongoing challenges of distributed hardware maintenance, and a now dated and bespoke control framework and on-node OS.  EdgeNet is the future of PlanetLab Europe, and it offers enormous advantages to both users and sites: users will be able to build applications on their laptops and deploy container pods as applications, and sites will be able to run PlanetLab Europe in VMs.

EdgeNet will permit PlanetLab Europe to offer a broader range of policies (both inside the host firewall and outside of it, as opposed to inside-the-firewall only today), and will permit the deployment of PlanetLab Europe/EdgeNet on a broad range of devices, including Raspberry PIs on up.  

The overall goal of PlanetLab Europe is to be the edge for arbitrary central X,  where X can be GÃšant, Measurement Lab (MLab), RIPE Atlas, or the Data Transparency Lab.  This is already accomplished with DTL and talks are ongoing with MLab.}

\item
\presentation{Ingrid Moerman (imec -- UGent, Belgium)}
{Virtualisation of different wireless networks sharing the same infrastructure}
{https://drive.google.com/file/d/1l_H6ZzXeAAR331U6raOQtGI_qjh5VErA/view}
{This presentation has shown some results, that have been showcased in the H2020 ORCA project (\url{http://www.orca-project.eu/}). Those showcases illustrate how wireless networks can be softwarized, virtualized and orchestrated and how this leads to more efficient use of spectrum and radio infrastructure, or improved QoS guarantees.
The following showcases have been presented:
\begin{enumerate}
\item imec's radio virtualization architecture enabling flexible PHY, flexible MAC and flexible networking and implemented on a Zynq-based System-on-Chip (SoC) SDR platform 
\item imec's architecture has been applied for tuning latency by changing sampling rate/bandwidth for IEEE 802.15.4 technology. Simple tweaks of the low cost sensor network standard can lead to more reliable or extended range communication (by reducing sampling rate) or lower latency communication (by increasing sampling rate)
\item It has been demonstrated that imec's architecture can support 11 concurrent radios on a single SDR board (2 x W-Fi, 8 x IEEE 802.15.4, 1 BLE)
Radio virtualization techniques also allow for eNB infrastructure sharing, where each operator can use its own spectrum on a shared radio infrastructure with only 1 antenna
\item In a joint experiment by imec and TCD, efficient spectrum sharing between coexisting networks in ISM bands has been illustrated by using radio slicing in combination with deep learning for technology recognition and traffic behaviour analysis
\end{enumerate}
}

\item
\presentation{Jerry Sobieski (NORDUNet, Denmark)}
{Untitled}
{http://indico.rnp.br/getFile.py/access?sessionId=5&resId=2&materialId=0&confId=260}
{}

\item
\presentation{Aki Nakao (University of Tokyo, Japan)}
{Untitled}
{http://indico.rnp.br/materialDisplay.py?sessionId=5&materialId=0&confId=260}
{This presentation explored applications of edge clouds in the IoT and as an adjunct to wireless networking.  Specific examples came from  image recognition: identifying objects and counting people from video on drones.  These applications typically require too much computation to be easily runnable at an edge device, but need   high bandwidth low latency connections to edge compute devices to run the deep learning algorithms required.  Adaptive, intelligent networking is needed to classify  applications on the fly, permitting application-specific edge processing. 

Network-aided object recognition. If the drone has a small GPU, classification can only be done on a  small number of frames per second, which is bad for object recognition. So send the traffic down to the ground to get it processed.  But flying 100 drones multiplies this traffic by 100.   So the solution with an edge cloud is to split the deep neural network into two parts: send the low bandwidth information from between two internal layers instead of sending the entire video. Enhance this with compression layers before sending.}

\item
\presentation{Prasad Calyam (University of Missouri, USA)}
{EdgeNet potential for supportive Cognitive IoT Applications}
{https://drive.google.com/open?id=1Q3cM4EumXnIRFHtUao-pZ3coqqZygS2s}
{This  talk presented  opportunities and challenges to develop and scale ``Cognitive IoT Applications'' (CIoT) using a transnational infrastructure as a modern implementation of a distributed edge cloud. CIoT involves analysis of sensor data collected at multiple sites for e.g., homes of elders, next-generation transportation systems, and smart buildings. It referenced  recent work on a related EdgeNet vision document \cite{CISEFuture2018} and presented use cases from eldercare and next-generation transportation where we see challenges in handling CIoT with core cloud and edge cloud integration.}
\item
\presentation{Ken Lutz (University of California -- Berkeley, USA)}
{Enabling Edge Computing With Cryptographically Hardened Data Containers}
{https://drive.google.com/open?id=0B1VYUMkK38mhOGFMdUFtWXpQdWZpUnUxTzZWcnJrQ0c3YWFR}
{In today's world, storage and processing of information is highly centralized in data-centers that are far away from users at the edge. This disparity of where the information is produced/consumed vs. where it is stored/processed only slightly affects the applications of today, but it will be the limiting factor for the applications of tomorrow that need low-latency access to storage/computation. Cheap yet powerful computational resources at the edge present an opportunity to develop rich low-latency applications, and like many others, we believe that ``edge computing'' will be more and more relevant. Going beyond merely filtering, caching and preprocessing of data using on-board computing resources, we envision the proliferation of networked resources at the edge and elsewhere on the path to cloud data-centers. Such resources can come in the form of small boxes in homes; on-premise servers for small-businesses, factory floors, and corporate offices; pico data-centers of the order of a server rack placed near end-users and managed by existing utility providers or municipalities; etc.

This edge infrastructure is complementary to the cloud infrastructure.In today's world, storage and processing of information is highly centralized in data-centers that are far away from users at the edge. This disparity of where the information is produced/consumed vs. where it is stored/processed only slightly affects the applications of today, but it will be the limiting factor for the applications of tomorrow that need low-latency access to storage/computation. Cheap yet powerful computational resources at the edge present an opportunity to develop rich low-latency applications, and like many others, we believe that ``edge computing'' will be more and more relevant. Going beyond merely filtering, caching and preprocessing of data using on-board computing resources, we envision the proliferation of networked resources at the edge and elsewhere on the path to cloud data-centers. Such resources can come in the form of small boxes in homes; on-premise servers for small-businesses, factory floors, and corporate offices; pico data-centers of the order of a server rack placed near end-users and managed by existing utility providers or municipalities; etc. This edge infrastructure is complementary to the cloud infrastructure.
}
\item
\presentation{Glenn Ricart (US Ignite, USA)}
{EdgeNet and Smart Gigabit Communities}
{https://drive.google.com/open?id=1wVq2I5kmPb9WPxq60XafLdTqwVi4Laqf}
{The 26 Smart Gigabit Communities project is a well-connected ecosystem of communities who have taken an interest in applications and services forged through collaborations between civic, academic, and industry leaders to advance their smart and connected communities. This project is funded in part by the National Science Foundation and by the communities and their sponsors. Smart Gigabit Communities represents a sustainable movement of collaborating cities exploiting the network effect of sharing advanced applications and services and processes both within and between smart communities. This testbed is available to any researcher who would like to explore an idea that may have significant impact in improving "real life" (education, healthcare, transportation, public safety, etc.) through technological advances applied with social and economic sensitivity. Smart Gigabit Communities all have an advanced gigabit networking infrastructure that supports high-bandwidth and low-latency edge applications and services operating in synchrony with the rhythm of a vibrant community. Industry (including startup) and academic researchers are equally welcome.}
\end{itemize}

\subsection{Discussion Details}

\subsubsection{Ambitions and goals}

The largest idea was the opportunity to build the world's largest and most scalable edge cloud.  EdgeNet is the third generation of edge cloud, following the PlanetLab generation (PlanetLab and PlanetLab Europe) and the GENI generation, which includes large-scale successes such as VNode, G-Lab, FED4FIRE, SAVI, and GENI itself.  EdgeNet was informed by the failures and successes of these projects.  Its software-only infrastructure is designed for high scalability and ease of installation, maintenance, and use.  In contrast to hardware-dependent infrastructures, where both on-site and central maintenance were challenging issues, software-only infrastructures have drastically reduced maintenance costs and presents a far smaller burden to participating sites.  The opportunity to achieve an unprecedented  scale-out is immediate.  In many ways, EdgeNet is to the edge cloud what the world-wide web was to information transfer: an easy-to-install infrastructure that spreads by local action, and there is no reason why this couldn't have a footprint at least as large.

One immediate opportunity was the conversion of a substantial fraction of the current installed base of PlanetLab Europe to EdgeNet, and the maintenance of a permanent EdgeNet installation on the existing GENI and SAVI backbones.  Together, this would give  a permanent base of over 200 sites with little incremental cost, spread across the US, Europe, and Canada.  One caution is that SAVI is unmaintained and the GENI and PlanetLab Europe infrastructures are aging, so requesting VMs on the host institution's existing Clouds is a high priority; use of existing bespoke hardware is a transitional step.

Glenn Ricart of US Ignite proposed introducing EdgeNet to  the existing  Smart Gigabit Communities, to expand the footprint of EdgeNet, offer a very easy-to-maintain installation, utilize existing standard open-source tools such as the Kubernetes ecosystem.  The transition would take place first, by instantiating EdgeNet worker nodes on the existing SGC communities' GENI racks, then transitioning to general-purpose local clouds.

 EdgeNet has been informed by both the maintenance and use cases issues of previous infrastructures, and by modern use cases.  The most prominent of these is the NSF-VMWare Data at the Edge initiative.  The Swarm Lab team represented at the workshop by Ken Lutz will deploy the Global Data Plane on EdgeNet, as  a general, secure  data transport layer for IoT and other data.  The point was made that a GDP-like infrastructure effectively decoupled IoT sensors and actuators from analytical sources and sinks, dramatically lowering the attack surface for specific attacks.

\subsubsection{Consensus and Open Issues}

The utility of a modern distributed infrastructure as a research vehicle for all four regions was generally agreed.  A specific interest was in a library of services created for and deployed on this infrastructure, of which the GDP is the earliest and prototype example.  The GDP is an example of a service which could be very widely used, since it offers data replication, security, and efficient transport -- all challenges for anyone deploying a distributed or IoT application.

A distributed, opt-in infrastructure such as EdgeNet presents  significant new challenges, which are themselves research opportunities.  While there is no question that a viral, software-only infrastructure such as EdgeNet will be orders of magnitude cheaper to maintain than infrastructures with dedicated hardware, this is still not free, and scaling limits are likely: even the original web had scaling limits, which were only overcome with significant changes to the original HTTP protocol.  We can expect similar challenges in large-scale distributed clouds, particularly in scaling the existing management infrastructure (Kubernetes), dealing with high churn as sites join, leave, and then re-enter the infrastructure, and migration of service points-of-presence in response to these challenges.

\subsubsection{Potential collaborations}

EdgeNet is already a deep collaboration between the US, EU, and Japan, remarkably as only the US side has to date been funded and that very sparingly.  Project participants, who meet weekly and maintain the existing EdgeNet infrastructure, are:
\newline

\begin{tabular}{c|c}
Timur Friedman & Sorbonne \\
Ciro Scognamiglio & PlanetLab Europe \\
Justin Cappos  & NYU Tandon \\
Albert Rafetseder & NYU Tandon and University of Vienna \\
Glenn Ricart  & US Ignite \\
Ada Gavrilovska & Georgia Tech \\
John Kubiatowicz & UC Berkeley \\  
Hausi Muller & University of Victoria \\
Eric Allman & UC Berkeley \\ 
Kenneth Lutz & UC Berkeley \\ 
Matt Hemmings & US Ignite \\
K\'evin Vermeulen & Sorbonne \\
Burim Ljuma & Sorbonne \\
Ketan Bhardwaj  & Georgia Tech \\
Berat Senel & Sorbonne \\
Peter Frech & Sorbonne \\
Rick McGeer & US Ignite \\
Aki Nakao & University of Tokyo
\vspace{4mm}
\end{tabular}
with significant participation from all areas.  The collaboration, begun at GEFI 2017, should be deepened and extended.  Unlike previous international collaborations in this space, which were either largely pursued independently with frequent collaboration, or were initiated by researchers in one area, this collaboration is a true worldwide collaboration with active participation from all areas from the project's inception.   Currently, researchers from the EU and US work on the same code base and make active contributions on an ongoing basis.

Further collaborations were identified, particularly in two areas: IoT analytics and secure data transport of IoT data.  Specific collaborations were identified between   Johann Marquez-Barja of the University of Antwerp  the Swarm Lab of UC Berkeley, represented here by Ken Lutz, and Smart Gigabit Communities (Glenn Ricart)

Jerry Sobieski of Nordunet suggested the deployment of Edgenet nodes on G\'EANT and Nordunet for distributed big science applications, incorporating the ITERATE radio-astronomy proposal of Rick McGeer, Ilya Baldin of RENCI, Jim Cordes of Cornell and Maura McLaughlin of West Virigina University.

A number of external collaborations were suggested, notably with the Measurement Lab project of Google and the New America Foundation.  This collaboration, initiated by Timur Friedman of the EU and Rick McGeer of US Ignite, involves deploying last-mile Measurement Lab clients on EdgeNet nodes.

\section{Federated Testbed of Elastic Optical Networks}
Session summary was prepared by Malathi Veeraraghavan (University of Virginia, USA),
Naoaki Yamanaka (Keio University, Japan),
and Kristin Rauschenbach (Notchway Solutions, USA).

\subsection{Session purpose}

\paragraph{Background} Optical networks have traditionally been used to interconnect IP routers. The significant growth in capacity, driven, for example by application and end-system trends like the cloud, big data, wireless access growth and Internet of Things (IoT), can no longer be supported by simply adding more routers.  More efficient and flexible bandwidth allocation methods and more scalable optical multiplexing and switching approaches are needed.  In addition to higher capacity, optical systems offer reduced operational costs compared to electrical systems, and especially lowering energy consumption and cooling costs.  Fixed-grid WDM Reconfigurable Optical Add/Drop Multiplexer (ROADM) designs have evolved to Colorless Directionless Contentionless (CDC) ROADMs, which offer considerably greater flexibility in selecting wavelengths and routes when provisioning lightpaths. These CDC ROADMs are built using optical couplers/splitters, wavelength splitters based on arrayed wave guides, and wavelength-selective switches. Photonic switches are used to build larger-scale optical crossconnects. Combined with tunable lasers and burst-mode receivers, these CDC ROADMs and optical crossconnects are used in dynamic optical networks.

Recent advances in FlexiGrid have led to further flexibility, enabling Elastic Optical Networks (EONs). FlexiGrid brings several advantages: support for 400 Gbps and 1 Tb/s (using superchannels); ability to support circuits with different bit rates; smaller channel spacing made possible with coherent detection; ability to tradeoff reach vs. spectral efficiency; and dynamic networking \cite{6146481}. Key components/systems in EONs include FlexiGrid CDC-ROADMs and elastic-rate (bandwidth variable) transponders/transceivers. A tutorial on EONs \cite{6824237} illustrates the savings in spectrum and transponder costs enabled by FlexiGrid. This paper also shows how transparent reach of optical signals can be traded off with spectral efficiency by leveraging the flexibility offered in EONs to select the modulation scheme, symbol rate, ratio of FEC to payload, inter-subcarrier spacing within a superchannel, and inter-superchannel spacing (guard bands). An important advance for connecting Ethernet interfaces to EONs was reported in an ECOC paper \cite{6964121}. Such an interface is critical to integrating EONs into the present-day Internet. This and other recent research and commercial activities demonstrate the importance of development across network layers.  This includes accessible control-plane interfaces that extend down to the physical layer to expand functionality, hasten adoption of innovative components, and allow applications to better utilize EON's inherent flexibility.

\paragraph{Session purpose and goals}
The purpose of this session is to discuss the feasibility of deploying a federated testbed of Elastic Optical Networks
(EONs) in three countries: USA, Japan and China. The goals for the presentations-part of the session were to have speakers describe best practices from previous successful testbed projects as well as to address risks, such as (i) community risks: are there small communities of
researchers who do not see value in a broad testbed; (ii) relevance risks: what types of optical networking
researchers could use the testbed; and (iii) technology risks: how do we avoid rapid obsolescence? \cite{Elliott2016-UVA-Talk}. The goals for the discussion period of the session were to elicit questions and answers on the main challenges involved in deploying such an EON testbed.

EON technologies offer the possibility of creating an interesting new type of Internet. With its potential for high-rate, low-latency connectivity, a new set of applications can emerge. EON components/systems are currently expensive, though third-party vendors are rapidly decreasing the costs of transceivers, transponders and other components. Still the cost of creating optical testbeds is high, and the number of researchers engaged in advancing control-plane/management-plane solutions, and applications, for these networks is currently small but could grow with a testbed. An international collaboration with shared login access of multiple testbeds, which could potentially be separate initially and later interconnected via NSF IRNC links, could hence lead to better ROI.

\paragraph{Research problems}
Optical physical-layer researchers can experiment with new components, line systems, multiplexers and switches. Subsystem researchers can explore novel system configurations.  System researchers can explore new control methods and applications.
Network control and management, and application-layer researchers, can access all layers, and evolve their work quickly as underlying technologies mature and change.  Security research can be conducted at all layers of the stack, physical through application, and across layers.

Consider the following examples. Scalable inter-domain routing solutions are required to support advance reservations for circuits without requiring autonomous systems to share their topology. Extensions to IETF Path Computation Engine Protocol (PCEP) for advance reservation have been proposed but need to be implemented and evaluated at scale. Management-plane problems include fault management, configuration management, and inter-domain performance monitoring. Security should be built into the protocols designed for EONs from the very start to avoid problems faced in today's Internet. Operational challenges, such as certificate expirations, troubleshooting failures, and administrator errors, require innovative solutions. A sizeable community of networking researchers work on the above-listed problems, as well as on network-design optimization problems. These researchers primarily use simulations to evaluate their work. The availability of an EON testbed would greatly improve the quality of the solutions proposed for the above-listed problems.

Innovative solutions are required to integrate these EONs with their new applications into the current Internet. Multi-layer, multi-vendor, and multi-domain integration challenges need to be solved.

\subsection{Presentations}
This set of presentations includes three talks on optical testbeds, and four talks on applications for elastic optical networks.
\begin{itemize}

\item
\presentation
{Kristin Rauschenbach (Notchway Solutions, LLC, USA)}
{Optical Whitebox}
{http://indico.rnp.br/getFile.py/access?sessionId=2&resId=3&materialId=0&confId=260}
{What features of an ``optical white box'' would drive a successful international
federated testbed of elastic optical networks? Using an ``optical white
box'' concept to guide the effort can help mitigate the three types of identified risks: community risks, relevance risks, and technology risks.
Both at the physical layer and at the higher network layers, it is clear that a carefully constructed ``white box'' concept can help
provide sufficient cohesion to a large federated research effort to enable the kind
of scale that gives high impact without over constraining the novelty and technology
refresh needed to ensure the research remains vibrant and relevant at multiple
layers of the stack. This presentation highlighted some key design tenants for an optical white box
including: size/scale considerations, modular construction to manage cost, use of
open optical interfaces that accommodate a variety of emerging devices, accessible
programmable control plane, and replacement of the proverbial ``killer app'' with a
set of general, yet aggressive, performance goals that support heterogeneous use
cases.
A community effort to better define, or begin design of, such a white box could not
only help to mitigate the risks identified for a federated network testbed, but also
generate important vertically-integrated technology innovations that are so critical
to a successful future for elastic optical networking.}

\item
\presentation
{Andrea Fumagalli, (The University of Texas at Dallas, USA)}
{International Collaborations Spearheaded by PROnet: a Programmable Optical Network Prototype}
{http://indico.rnp.br/getFile.py/access?sessionId=2&resId=2&materialId=0&confId=260}
{PROnet is a Programmable Optical Research and Education Network deployed at and around the University
of Texas at Dallas campus. Initially funded by the NSF CC* initiative, the PROnet prototype has attracted
a number of international partners from both academia and industry. Each partner has offered at least one
unique technical contribution to the PROnet prototype, which would not have been there otherwise. This
presentation described the nature and outcomes of these partnerships, focusing on what has
worked and what has not. The following PROnet milestones have either been achieved or are in the
process of being pursued: integration of open source and proprietary software modules to achieve a highly
reliable SDN-controlled two-layer Ethernet over DWDM network, integration of a Quality of Transmission
Estimator (QoT-E) module provided by the research team of Prof. Vittorio Curri (Politecnico di Torino -
EU), integration of a Cloud Radio Access Network (C-RAN) solution with built-in reliability of both
backhaul and fronthaul mobile network segments achieved in collaboration with the research team of Prof.
Luca Valcarenghi (Scuola Sant'Anna -- EU), integration of Virtual Machine (VM) live migration
functionalities driven by an application provided by the research team of Prof. Naoaki Yamanaka (Keio
University - Japan), and integration of an optical space switch SDN controller provided by the research
team of Prof. Gustavo Sousa Pavani (Universidade Federal do ABC -- Brazil). The talk will also discuss
how some of the PROnet prototype outcomes align with industry-led efforts like the Telecom Infra Project
(TIP) and Open ROADM.}

\item
\presentation
{Joe Mambretti (Northwestern University, USA)}
{Federated Elastic Optical Networking Research Testbeds}
{http://indico.rnp.br/getFile.py/access?sessionId=2&resId=1&materialId=0&confId=260}
{In today's fast changing dynamic environment, static optical networks that requiring significant manual operational
efforts are no longer adequate, especially with regard to the need for capacity increases, and flexibility
and programmability of the optical infrastructure. Currently, International Center for Advanced Internet Research (iCAIR) supports over 25
networking testbeds, including a number of national and international optical testbeds.
Optical networking research topics being addressed by iCAIR include optical disaggregation
based on Software Defined Networking (SDN), optical service integration with Software
Defined Exchanges (SDXs), SDN integration with bit-rate coherent optics using tunable, flexible
grid reconfigurable photonic layers, open APIs, P4 programmability with optical channels,
optical network services and fabrics, and AI/ML/DL potentials for optical services optimization,
configuration, reliability, and general operations. These emerging techniques provide a fluidity to
optical networks to make them pliable, i.e., elastic, including dynamic real-time changing
resources. For SC18, iCAIR partnered with SCInet to provision a national optical
fabric that would support over 40 national and international 100 Gbps demonstrations.}

\item
\presentation
{Weiqiang Sun (Shanghai Jiao Tong University, China)}
{Large Scale Trial of Hybrid Packet/Circuit Switched Technology on a Cross-country Testbed}
{http://indico.rnp.br/getFile.py/access?sessionId=2&resId=4&materialId=0&confId=260}
{In the current Internet, large flows co-exist with small flows and often consume a significant
proportion of the network bandwidth for a considerably long period of time. This imposes a
challenge on network resource management, in both intra-datacenter and inter-datacenter
network. In the recent past, many research efforts have been undertaken exploring
the use of hybrid packet/circuit switched technology in intra-datacenters networks (DCNs) and those
efforts has provided interesting insights into the design of future DCNs. However, due to
daunting complexity of managing bandwidth across a large-scale public network, little work
has been done on the use of hybrid switching technology over such a scale. This presentation
summarized the state of art of hybrid switching in DCN: progress made, lessons
learned and challenges ahead. A resource allocation
framework called BLOC -- Blocking LOss Curve for use in
hybrid large-scale networks was discussed.
Indeed the use of hybrid switching in large-scale networks is much more complex than that in
DCNs. Experiments over a cross-country testbed will be highly valuable in determining the
real difficulties in such an endeavor. That said, it would be of great value to bring together the
community and discuss how such experiments could be conducted, and what could be achieved.}

\item
\presentation
{Prasad Calyam (University of Missouri, USA)}
{Leveraging GreyFiber for Geo-Distributed Latency-Sensitive Computer Vision Analytics}
{http://indico.rnp.br/getFile.py/access?sessionId=2&resId=0&materialId=0&confId=260}
{This talk presented opportunities and challenges to perform ``Geo-Distributed Latency-Sensitive
Computer Vision Analytics (CVA)'' on a federated testbed of
elastic optical networks. CVA involves analysis of video/image
data collected at multiple sites for disaster response
coordination, UAV-supported agriculture, wide-area
surveillance, and other applications. Particularly, the talk presented results from the
recent work on
GreyFiber \cite{durairajan2018greyfiber}, and the use
case of real-time video analytics, which is described as a killer app for edge computing by Microsoft Research
\cite{8057318}. The talk describes the challenges in handling the large data volumes and analysis
needs for users/operators relying on CVA.}

\item
\presentation 
{Naoaki Yamanaka (Keio University, Japan)}
{5G and Elastic Optical Edge Network Testbed for Autonomous Driving Vehicle}
{http://indico.rnp.br/getFile.py/access?sessionId=2&resId=6&materialId=0&confId=260}
{Everything is being connected to the Cloud and Internet of Things, and network robots with big data analysis
are creating important applications and services. The cloud network architecture is moving towards mega-cloud data
centers (DCs) provided by companies such as Amazon and Google, in combination with distributed small DCs or edge
clouds. While the traditional restrictions imposed by distance and bandwidth are being overcome by the
development of advanced elastic optical networks that offers flexible slices and high-bandwidth, modern
applications impose more complex performance and quality of service requirements in terms of processing power,
response time, and data amount. The rise in cloud performance must be matched by improvements in network
performance. An application-triggered cloud network architecture based on huge-bandwidth, logical local
mesh, elastic optical interconnections was proposed \cite{8410212}.
This talk presented concepts of how to use virtual machine migration between
edge clouds, as well as between edge clouds and center clouds. An example application is to support
Autonomous Driving Vehicles (AD-cars). The talk presented architectures using technologies that can realize energy-efficient
and high-performance cloud service. In addition, a demonstration system of an AD-car control by an edge computer was
described. To guarantee the response time at 10ms, VM migration techniques were used to follow vehicle movements.
A flexible and programmable router, based on a resource pool architecture, was used in this demonstration system.
A number of cloud-based applications are expected to offer advanced services to vehicles. Some of the applications will
be running on VMs on edge computers to have short network round trip latency with the vehicle. As the vehicle moves to
other physical locations, the VM will be need to  be ``live migrated'' along with its assigned vehicle, without interruption of
service. A collection of numerous Edge facilities next to the road infrastructure is assumed.
The talk reported demonstration results of dynamic VM migration using programmable edge node and optical
elastic networks under the control of an orchestrator. In the K2-campus of Keio University, an AD-car control
network testbed with 5G and elastic optical mesh networks has been deployed.}

\item
\presentation
{Malathi Veeraraghavan (University of Virginia, USA)}
{Science Elastic Optical Inter-Network (SEOIN)}
{http://indico.rnp.br/getFile.py/access?sessionId=2&resId=5&materialId=0&confId=260}
{Scientific-computing applications have high capacity and low-latency requirements that can be met with optical
 networks.  The overall vision presented in this talk is to realize a global-scale Science Elastic Optical Inter-Network (SEOIN) that offers high-speed end-to-end rate-guaranteed dynamic Layer-1 (L1) circuits. The vision leverages two trends: (i) elastic optical networks enabled by FlexiGrid, and (ii) software-defined networking (SDN). The term ''inter-network'' is used to emphasize that the new protocols and algorithms
implemented in SDN controllers should be designed to support a multi-domain (multi-organization) deployment. Integration challenges,
control-plane and management-plane challenges, and adoption challenges were described in this talk.}

\end{itemize}

\subsection{Discussion Details}

\subsubsection{Ambitions and goals}

There is a bottom-up driver for optical network research to reduce the complexity introduced by the impressive
set of recent innovations in optical technologies, such as tunable lasers, coherent transponders, wavelength selective switches,
ROADMs, elastic grid technologies, silicon photonic switches \cite{Shen:18}, etc. Disaggregated systems, and correspondingly orchestrators needed for interconnecting the disaggregated components,
are two key trends that have enabled new research problems. When a system is disaggregated, it creates the potential for purchasing components from different vendors, which increases competition and drives down costs.

Furthermore, there is a top-down driver for optical network research driven by new applications, such
as intra- and inter-datacenter networking \cite{8293990}, metro and wide-area networking, Passive Optical Networks (PONs), RF-optical integration, Cloud Radio Access Networks (CRAN), smart city
applications that require low-latency communications between edge clouds and end devices, such as autonomous vehicles (significant amounts of data will be generated from the sensors on such cars),
geo-aware resilient computer vision analytics, especially with live video from disaster areas, and
scientific-research driven applications such as stream processing with online steering of remote applications and
bulk-data transfers in genomics, connectomics and high-energy physics.

Finally, research opportunities arise from advances in Artificial Intelligence (AI), Machine Learning (ML) and Deep Learning (DL). Specifically, these methods can be used to reduce the high costs of optical network management and operations.

To support researchers addressing these optical network problems, an experimental federated elastic optical testbed
would be most useful. There are many challenges in building such a testbed. \emph{We identified two approaches}: (i) design and implement
an optical whitebox, and deploy in many locations, and (ii) deploy optical testbeds using off-the-shelf equipment. In both cases,
standards are required for ``open'' interfaces, ports, and control points, to (i) support enough flexibility for researchers to
``plug and play'' with novel components, node types, control schemes and applications, and (ii) to have sufficient compatibility to
enable large-scale infrastructure deployment.

Commercial efforts to disaggregate optical transport and enable tighter electrical/optical
packet integration methods are being pursued by the Open Compute Project (OCP), Telecom Infrastructure Project (TIP), and
vendors such as Lumentum, ADVA and Edge-core \cite{Robinson:18}. Whitebox solutions such as Voyager and Cassini were
presented at recent TIP summits. These commercial efforts should be carefully 
leveraged in our testbeds.

The following ``big ideas'' were generated:
\begin{enumerate}
\item Build simple white boxes that use components from companies such as Finnisar, and exclude many of the archaic protocols found in optical systems designed for telcos. The concept of taking incremental steps, and gradually adding complexity, was proposed.
\item We need to coordinate activities across the different layers in order for our academic research to have bigger impact.
For example, researchers working on devices with new modulation schemes could work
together with control-plane researchers to develop protocols that incorporate the use of the new modulation
schemes. We decided that it was imperative to include researchers from multiple disciplines in the testbed planning effort. Another example was provided. Significant engineering effort was required in an academic lab to build $3 \times 3$ switches with ns-switching speeds, and to integrate tunable lasers with optical switching. But these components should be used by other researchers in their experimental plans in order to advance these new systems. This requires coordination between research groups, and definition of common interfaces (APIs) for components.
\item Use AI, deep learning and machine learning for improved optical network administration.
\item Explore the integration of big-data scientific applications on optical testbeds with end-to-end 100G flows. Examples include Multicore-Aware Data Transfer Middleware (MDTM)\footnote{http://mdtm.fnal.gov/index.html} computational astrophysics, genomics, weather exploration, activities at NRL and NASA's Goddard Space Flight Center, and biomedical informatics.
\item Recruit academics to run educational activities on the optical tested. Educational use was an important growth driver for
the GENI user base.
\end{enumerate}

Challenges include the following:
\begin{enumerate}
\item How we can design testbeds that simultaneously support researchers who work on problems in the
physical (optical) layer, and develop components and subsystems for optical line systems, transmission systems and switching systems, as well as
researchers who solve control-plane and management-plane problems, or experiment with network architectures, protocols, and applications?
There is a time-length difference between research at the physical layer and at higher layers. The former usually involves hardware design, which takes longer than the software prototyping required to test and evaluate solutions for higher-layer problems.
\item Optical systems are typically expensive. Recent advances though have lowered costs. For example, 
the international Center for Advanced Internet Research at Northwestern University, the StarLight
International/National communications Exchange Facility consortium and Ciena collaborated to create a testbed
based on private optical fiber between the StarLight facility in Chicago and the Ciena Research Lab in Ottawa.
Ciena Waveservers were placed on either side of this 1440-km path. Using this testbed, three 100 Gbps lightpaths
were bonded in a superchannel to achieve 300 Gbps E2E transmission. The discounted cost of the Waveservers including optics was \$50K per site, \$100K total. Also, Voyager and Cassini whiteboxes were announced at costs of about \$75K.
\item Can all components, subsystems and systems be made remote controllable? If not, manual remote hands-and-eyes operations at testbed locations can slow down research projects. If each new experiment requires manual adjustments/upgrades, testbed administrative (HR) support could become a bottleneck.
\item How do we to avoid technology lock-in when creating testbeds with vendor supplied equipment?
\item How do we monitor the testbeds to verify that services are working, detect failed services, and run automated
fault analysis and recovery methods?
\item Should the initial set of testbeds be deployed at university laboratories without interconnections? A second step would be to add long-distance point-to-point optical circuits to interconnect separate testbeds. A final step would be to deploy optical ROADMs and switches at PoPs (e.g., Internet2, ESnet, Global Lambda Integrated Facility Open
Lambda Exchanges (GOLEs), including the AutoGOLEs, JGN and CERNET PoPs) and further, allow for researchers to connect and test new components, subsystems and systems at these PoPs.
\item Integration methods that enable applications running on hosts and other end devices to leverage the high capacity and low-latency of optical circuits are required. For example, can  P4 concepts \cite{Bosshart:2014:PPP:2656877.2656890} be used to connect optical channels to file systems? What types of transport and link-layer protocols are required on the data-plane when the underlying network paths are optical lightpaths instead of IP-routed paths?
\end{enumerate}

The following next steps were identified:
\begin{enumerate}
\item Work with Tier-2 vendors. Attendees mentioned the use of this strategy to then later gain the attention of Tier-1 vendors.
\item Understand ongoing efforts in Open Network Automation Platform (ONAP),\footnote{https://www.onap.org/} Telecom Infra Project
(TIP),\footnote{https://telecominfraproject.com/} and Open ROADM.\footnote{http://openroadm.org/home.html}  For example, we should study APIs such as Open ROADM Multi-Source Agreement (MSA), which covers both optical interoperability as well as YANG data models.\footnote{https://wiki.onosproject.org/display/ONOS/Open+ROADM+MSA}  In addition, MEICAN is a set of tools that was developed by the Brazilian National Research and Education Network (RNP).  MEICAN is a tool wrapper for the Network Service Interface (NSI 2.0), which is an architectural standard developed by the Open Grid Forum as an API for network controllers. NSI 2.0  is an East West protocol for dynamic multi-domain L2 provisioning that can also be used for dynamic L1 lightpath provisioning, and is being implemented in a number of production R\&E networks and exchange points.
\item Have academics from our group attend TIP conference calls.\footnote{https://telecominfraproject.com/open-optical-packet-transport/} This tactic has been useful in the radio domain.
    \item Start establishing partnerships with system vendors. While they offer Yang, TL1, CORBA plugins, we cannot control power level, amplifier gain, etc. We need to define APIs to be able to control the power levels of channels.
\item Explore ideas for a post-GENI distributed network research infrastructure presented in \cite{CISEFuture2018}, and relate this proposal to our goals for creating a federated elastic optical network testbed. Key ideas are to have a programmable core, storage and compute in core and at the edge, and to link various facilities (GENI, NSF Clouds, PAWR, large supercomputers, scientific instruments and campuses).
\item Create a Research Coordination Network (RCN) for elastic optical networks, and attract industry participation.
\item Engage with CpQD\footnote{https://www.cpqd.com.br/en/} and Padtec\footnote{http://www.padtec.com.br/en/} from Brazil. Padtec is a spinoff from CPQD. These companies have significant experience with packaging. CpQD deployed a 5-node FlexiGrid testbed using their own equipment.
\item For the whitebox development effort, draw a Venn diagram of what we want to build. Limit functionality, and use spiral development methods.
\item Consider the NSI standard for optical circuit provisioning. There is a strong connection abstraction, and NSI is seeing broad adoption in Layer-3 and Layer-2 SDNs.
\end{enumerate}

\subsubsection{Consensus and Open Issues}
The group definitely reached consensus about needing such elastic optical testbeds in order to enable our research to have higher impact.  Elastic optical networks are a key to addressing both high-speed and flexibility, wich is important because different applications and service and network architectures demand different operational speeds.  In addition, speed and flexibility are determined not only by the control system/protocol but also device technology. This drives a need for vertical collaboration among device researchers, system designers and control-plane engineers. With technologies progressing dynamically, researchers from different disciplines must work closely.

We also agreed on the need to pursue the development of optical white boxes, and the deployment of testbeds with vendor
supplied systems. The need for open APIs was widely supported. The need to support both physical-layer and higher-layer optical network research had wide consensus.

Open issues are primarily the challenges listed in the previous section.

\subsubsection{Potential collaborations}
Current collaborations include the following: (i) The University of Texas at Dallas (Andrea Fumagalli), Keio University (Naoaki Yamanaka),
Scuola Sant'Anna - EU (Luca Valcarenghi), Universidade Federal do ABC -- Brazil (Gustavo Sousa Pavani), Politecnico di Torino -
EU (Vittorio Curri); and (ii) Shanghai Jiao Tong University (Weiqiang Sun) and University of Virginia (Malathi Veeraraghavan).

Northwestern University (Joe Mambretti) has multiple optical network research collaborations.\footnote{For details, see Joe Mambretti's presentation, ``Federated Testbed of Elastic Optical Networks.''}

UTD (USA) and Keio University (Japan) plan testbed research for elastic optical network (with optical links, ROADMs and switches) orchestration to support multiple applications, including autonomous driving vehicles and distributed data centers. This testbed effort includes system vendors, a service provider and a test equipment provider.

We have identified other US research groups led by Keren Bergman (Columbia University),
Dan Kilper (University of Arizona), S. J. Ben Yoo (University of California, Davis), and
Rongqing Hui (University of Kansas) as users or providers of optical testbeds.  US researchers that can add (i) wireless perspective include Ivan Seskar (Rutgers) and Abhimanyo Gosain (Northeastern University), (ii) data center networks perspective include George Porter (UCSD), and (iii) infrastructure and applications perspective include Ilia Baldine and Paul Ruth (RENCI), and Jerry Sobeiski (NORDunet). From Japan, Akihiro Nakao (University of Tokyo) is interested in designing and prototyping optical whiteboxes, and contributing to the convergence of computer science and wireless/wired networks.  Researchers interested in optical testbeds include, from Japan, Prof. Eiji Oki, Kyoto Univ., Prof. Satoru Okamoto, Keio Univ., Prof Kohei Shiomoto, Tokyo Metropolitan Univ.,  and from China, Prof. Gangxiang Shen, Soochou University (shengx@suda.edu.cn), and Prof. Zuqin Zhu, University of Science and Technology of China (zqzhu@ieee.org).

\section{Reproducibility in Experimentation}

Session summary prepared by Lucas Nussbaum (Universit\'{e} de Lorraine, France) and Violet Syrotiuk (Arizona State University, USA).


\subsection{Session purpose}

Reproducibility is a fundamental part of the scientific method. 
It is different from \textit{repeatability} where researchers repeat their own experiment to verify their results, and \textit{replicability} where an independent group of researchers uses the original experimental set-up to verify results.
\textit{Reproducibility} consists of a replication study performed by an independent group of researchers using their own experimental set-up to confirm the results and conclusions of an earlier experiment. 

Today, there is a crisis in reproducibility across many scientific disciplines, including reproducing the empirical results from the engineered networks used by the international GEFI community. 
Symptoms of the reproducibility crisis include the selective reporting of results, poor analysis of results, non-standardized lab methods, unreported methods, poor experimental design, unavailability of the original data, and insufficient peer review, among others.
Experimentation becomes robust when it is easily repeated. 
Hence this session proposes measures to address the reproducibility crisis including tools to support experimenters, including improved experimental design and statistical analysis, standardizing lab methods, and experimental repositories.
It is essential that the members of the GEFI community collaborate on this problem due to its ubiquity, and because otherwise the conclusions reported from the data collected are reduced to hearsay, reducing progress in our field.

Goals of the session were to discuss:
\begin{itemize}
	\item tools and services for improving experiment reproducibility (including, e.g., the use of notebooks for experiment control),
	\item data management practices and experimental reproducibility (i.e., provenance tracking, data preservation, etc.), and
	\item lessons learned in trying to achieve reproducibility in practice.
\end{itemize}


\subsection{Presentations}

The session included the following seven presentations.

\begin{itemize}

\item
\presentation
{Lucas Nussbaum (Universit\'{e} de Lorraine, France)}
{Experiment Data Management}
{http://indico.rnp.br/getFile.py/access?sessionId=3&resId=5&materialId=0&confId=260}
{The proper sharing of research artifacts is a requirement for reproducibility.
A \textit{data management plan} (DMP) is one aspect of reproducibility, and includes storing data \textit{during}, \textit{between}, and archiving data \textit{after}, experiments.
The de facto standard in archiving is GitHub, but data repositories require exploration.
For storage, Swift-based object stores are available, as are per-project or per-user NFS directories.
Grid'5000 has two recent developments in the area of experiment data management. 
It provides a disk reservation system that allows experimenters to keep their data on nodes between reservations.
It also provides some security improvements on the storage systems used for medium-term storage of experiment data.}


\item
\presentation
{Brecht Vermeulen (Ghent University -- imec, Belgium)}
{Experiment Reproducibility with jFed}
{http://indico.rnp.br/contributionDisplay.py?contribId=32&sessionId=3&confId=260}
{Building on existing formats such, as resource specifications (RSpecs) and tools such as jFed, an experiment specification (\textit{ESpec}) is defined for creating reproducible experiments.
An ESpec bundles the resources specifications, scripts, and data sources for the full and automatic deployment of an experiment. 
Examples of ESpecs include scheduling of jobs in the GPU lab, and testbed federation monitoring due to continuous integration of new testbed elements.}


\item
\presentation{Kate Keahey (Argonne National Laboratory and the University of Chicago, USA)}
{Repeatability as Side-Effect}
{http://indico.rnp.br/contributionDisplay.py?sessionId=3&contribId=33&confId=260}
{A  pr\'{e}cis in Chameleon represents information about user experiments collected as a side-effect of running an experiment.
It provides a record of the resources used in the experiment as well as an analysis of the results, allowing the experiment to be repeated potentially with variations. 
An experiment pr\'{e}cis is analogous to the Linux ``history'' command, reflecting the actions taken by the user  when interacting with the system.
It can be edited or processed to, e.g., simplify the workflow it represents, or streamed to a file and turned into a script repeating the actions.
Experiment pr\'{e}cis have also been integrated with Jupyter notebooks.
These implementation options allow experiments to be easily shared with others.}


\item
\presentation
{Jason Liu (Florida International University, USA)}
{Virtual Time Machine for Reproducibility of Network Emulation Experiments}
{http://indico.rnp.br/contributionDisplay.py?sessionId=3&contribId=34&confId=260}
{With the increasing presence of shared network testbeds and compute infrastructure, researchers are now able to instantiate the same network experiment environment repeatedly.
However, it can be difficult to acquire a large number resources for prolonged experiments with large-scale networks. 
Emulation is one solution to this problem and it, together with the virtual time machine to provide better timing fidelity, can contribute to  reproducibility in large-scale network experiments.}


\item 
\presentation
{Prasad Calyam (University of Missouri, USA)}
{Custom Template Recommenders for Re-use/Re-purpose of Data-intensive Applications using Federated Resources}
{http://indico.rnp.br/contributionDisplay.py?sessionId=3&contribId=35&confId=260}
{A goal of a custom template is to promote repeatable/reusable resource provisioning and service composition satisfying diverse user and application needs.
Storing such templates in a catalogue help with reuse and repurposing workflows to other applications.
Such a catalogue has been developed to meet user needs in the CyNeuro Gateway, a web portal for software and cyber automation tools in neuroscience to address the emerging needs involving big data.
These needs include providing a reproducible computing environment, customizing pipelines and the ability to re-run analyses on the same data, quick data processing (using local or remote resources), and compelling visualizations to share with scientific community.}


\item
\presentation
{Graciela Perera (Northeastern Illinois University, USA)}
{Experimentation based Teaching Innovation using Testbeds}
{http://indico.rnp.br/contributionDisplay.py?sessionId=3&contribId=36&confId=260}
{Testbeds can be used by students to gain experience solving problems in a collaborative setting.
It allows students to connect concepts in networking with ``hands-on'' experience, and benefit from peer interaction.
Specifically, testbeds are used at NIU to showcase fundamental concepts in network security and distributed systems for students with highly diverse educational and cultural backgrounds.
In turn, instructors learn new approaches to teaching that can be adapted and applied to successive courses.}


\item 
\presentation
{Violet Syrotiuk (Arizona State University, USA)}
{Screening Experiments}
{http://indico.rnp.br/contributionDisplay.py?sessionId=3&contribId=37&confId=260}
{Complexity in engineered networks arises from their size, structure, operation, evolution over time, and the fact that humans are involved in their design and operation.
Before conducting experiments in such networks, a screening experiment should be conducted to identify factors (parameters) that significantly influence the responses (e.g., throughput, energy, delay, etc.) measured. 
Screening provides an objective confirmation that the factors being varied in the experiment are significant.
Classical techniques for screening such as fractional-factorial, D-optimal, and super-saturated designs are either too large or emphasize the screening of main effects only.
Locating arrays are a new screening design that grow only logarithmically in the number of factors, and can also identify interactions.
Using a screening experiment is an important first step in reproducing any experimental work.}

\end{itemize}


\subsection{Discussion Details}

\subsubsection{Ambitions and goals}


\begin{itemize}

\item Funding agencies in the USA and Europe (and Asia?) require proposals to include data management plans but what these must include  are very vague.

\item Should papers include a repository with data artifacts and/or an experiment specification such as a jFed ESpec or a Chameleon pr\'{e}cis?

\item Europe has now mandated that publications resulting from funded projects must appear in Open Access journals.
What are the implications to researchers, and to publishers?  
Does it restrict publication to those who have funding?

\end{itemize}


\subsubsection{Consensus and Open Issues}


\begin{itemize}

\item Our community could perhaps benefit from more expertise in experimental design and analysis of experimental results (e.g., statistics, data science, visualization).

\item Incorporating more training in these topics and associated tools in computer science and engineering disciplines is advised.

\end{itemize}


\subsubsection{Potential collaborations}


\begin{itemize}

\item Violet Syrotiuk seeks a collaboration involving large-scale experimentation involving heterogenous testbeds (e.g., wireless and optical).  She is interested to investigate the benefits of screening experiments to focus experimentation in such settings.

\end{itemize}


\section{Detailed Information}
\label{sec:detailed-information}

This section presents detailed workshop information. Additional information is available
from the workshop web page: \url{http://indico.rnp.br/conferenceDisplay.py?confId=260}.

\subsection{Workshop Participants}

Participants were selected via an open call process, based on position statements
submitted in advance and reviewed by the workshop organizers. The following
people participated in GEFI 2018. Each participant had the opportunity to contribute
to any sessions of interest, both through formal presentations and in group discussion.

\begin{itemize}
	\item Mark Berman, \textit{GENI Project Office, Raytheon BBN Technologies} (USA)
	\item Prasad Calyam, \textit{University of Missouri} (USA)
	\item Leandro Ciuffo, \textit{RNP} (Brazil)
	\item Ping Du, \textit{University of Tokyo} (Japan)
	\item Masatoshi Enomoto, \textit{NICT} (Japan)
	\item Timur Friedman, \textit{Sorbonne Universit\'e} (France)
	\item Andrea Fumagalli,* \textit{University of Texas, Dallas} (USA) 
	\item Ada Gavrilovska, \textit{Georgia Institute of Technology} (USA)
	\item Abhimanyu Gosain, \textit{Northeastern University} (USA)
	\item Hiroaki Harai, \textit{NICT} (Japan)
	\item Eiji Kawai, \textit{NICT} (Japan)
	\item Kate Keahey, \textit{University of Chicago} (USA)
	\item Dongkyun Kim, \textit{Korea Institute of Science and Technology Information} (Korea)
	\item Jason Liu, \textit{Florida International University} (USA)
	\item Kenneth Lutz, \textit{University of California, Berkeley} (USA)
	\item Iara Machado, \textit{RNP} (Brazil)
	\item Joe Mambretti, \textit{Northwestern University} (USA)
	\item Anirban Mandal, \textit{RENCI} (USA)
	\item Cesar Marcondes, \textit{Aeronautics Institute of Technology} (Brazil)
	\item Johann Marquez-Barja, \textit{University of Antwerp} (Belgium)
	\item Rick McGeer, \textit{US Ignite} (USA)
	\item Deep Medhi, \textit{NSF} (USA)
	\item Toshiyuki  Miyachi, \textit{NICT} (Japan)
	\item Ingrid Moerman, \textit{IMEC, Ghent University} (Belgium)
	\item Akihiro Nakao, \textit{University of Tokyo} (Japan)
	\item Lucas Nussbaum, \textit{University of Lorraine} (France)
	\item Graciela Perera, \textit{Northeastern Illinois University} (USA)
	\item Kristin Rauschenbach, \textit{Notchway Solutions} (USA)
	\item Srivatsan Ravi, \textit{University of Southern California} (USA)
	\item Jos\'e Rezende, \textit{RNP} (Brazil)
	\item Glenn Ricart, \textit{US Ignite} (USA)
	\item Marco Ruffini, \textit{Trinity College Dublin} (Ireland)
	\item Paul Ruth, \textit{RENCI} (USA)
	\item Stephen Schwab, \textit{University of Southern California} (USA)
	\item Ivan Seskar, \textit{Rutgers University} (USA)
	\item Jerry Sobieski, \textit{NORDUnet} (Denmark)
	\item Weiqiang Sun, \textit{Shanghai Jiao Tong University} (China)
	\item Violet Syrotiuk, \textit{Arizona State University} (USA)
	\item Yuuichi Teranishi, \textit{NICT} (Japan)
	\item Peter Van Daele, \textit{IMEC, Ghent University} (Belgium)
	\item Malathi Veeraraghavan, \textit{University of Virginia} (USA)
	\item Brecht Vermeulen, \textit{IMEC, Ghent University} (Belgium)
	\item Ann Von Lehman, \textit{NSF} (USA)
	\item Naoaki Yamanaka, \textit{Keio University} (Japan)
\end{itemize}

* Indicates remote participation.

\subsection{Agenda}

GEFI 2018 was held over a two day period, October 25-26, 2018. The agenda is 
summarized below. The full agenda, including links to presentations, is available
from the workshop web site.\footnote{See \url{http://indico.rnp.br/conferenceTimeTable.py?confId=260}.}

\begin{figure}[H]
\frame{\includegraphics[width=\columnwidth]{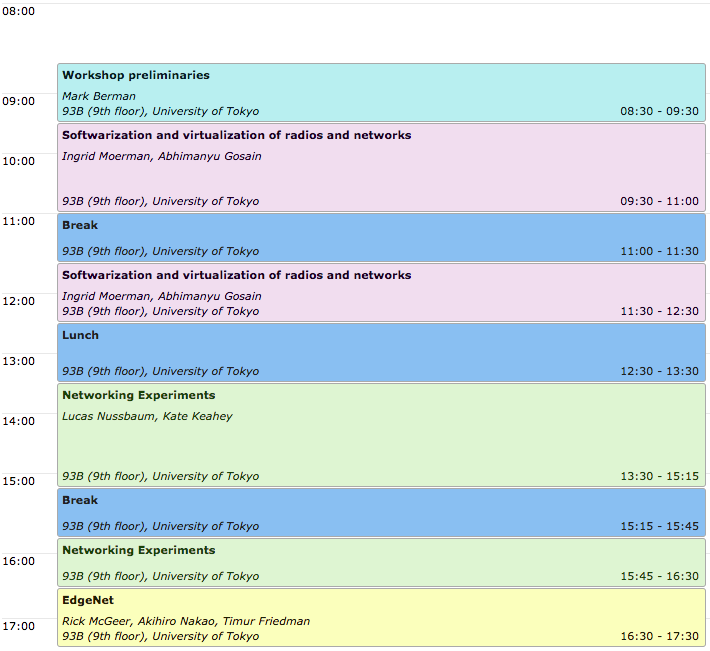}}
\caption{Workshop day one agenda.}
\label{fig:day-one-agenda}
\end{figure}

\begin{figure}[H]
\frame{\includegraphics[width=\columnwidth]{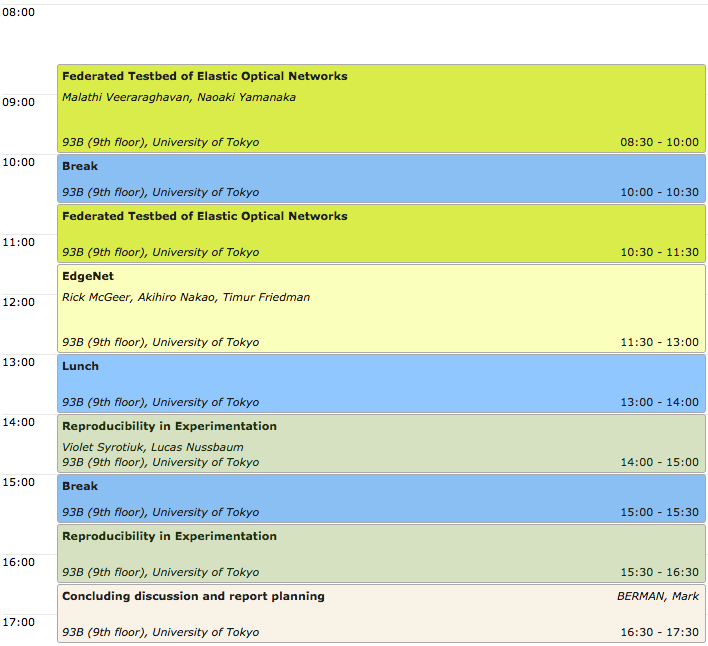}}
\caption{Workshop day two agenda.}
\label{fig:day-two-agenda}
\end{figure}

\subsection {Presentation Summary and Links}
\label{sec:presentation-summary}

The following list summarizes presentations at GEFI 2018, ordered by session,
and includes links to the information presented.
\insertPresentationSummary

\subsection {Testbeds Represented}

In keeping with GEFI's goals of encouraging testbed-supported research,
representatives of many research testbeds participated in the workshop.
See Table \ref{tab:testbeds} for a summary of testbeds represented.

\tablehead{%
\textbf{Testbed Name}&
\textbf{Sponsorship}&
\textbf{Description}
\\\hline
}
\tabletail{%
\hline
\multicolumn{3}{c}{\textbf{Table \ref{tab:testbeds}: Testbeds Represented (continues next page)}}\\
}
\tablelasttail{\hline}
\bottomcaption{Testbeds Represented}
\label{tab:testbeds}

\onecolumn

\begin{mpsupertabular*}{\textwidth}{p{0.15\textwidth}p{0.20\textwidth}p{0.60\textwidth}}

Global Environment for Network Innovations (GENI)&
U.S. National Science Foundation (NSF)&
GENI is a nationwide federated research testbed spanning the US. Its resources are sliced and deeply programmable. They include compute resources hosted in approximately sixty GENI racks, as well as thirteen LTE wireless base stations. These resources are hosted chiefly by universities and interconnected by a backbone network hosted by Internet2. GENI supports additional experimental resources through federation with a number of other US and international partner testbeds. GENI supports a wide variety of research and educational experiments in networking, distributed computing, and other scientific disciplines. See \url{http://geni.net}  or \cite{GENIBook2016} for more information.
\\\hline

Fed4FIRE &
European Commission sponsors the Fed4FIRE project, this is tooling and community for federation, not the testbed hardware itself.& 
See \url{https://www.fed4fire.eu/testbeds/} for the included testbeds. These testbeds are funded outside of the Fed4FIRE project.
\\\hline

w-iLab.t&
Local government Belgium (+advanced tooling sponsored by various Eur. Comm. projects )&
A testbed for wireless and IoT research and education with 300+ wireless nodes (in industrial, office and home environment), including mobile nodes. Technologies include 802.11a/b/g/n/ac, multiple types of software defined radios, LTE, a small shielded environment, 802.15.4, 868/434MHz radios.  More information at \url{https://doc.ilabt.imec.be/ilabt-documentation/}.
\\\hline

Virtual Wall& 
Local government Belgium (+advanced tooling sponsored by various Eur. Comm. projects)& 
A testbed for wireless and IoT research and education with 300+ wireless nodes (in industrial, office and home environment), including mobile nodes. Technologies include 802.11a/b/g/n/ac, multiple types of software defined radios, LTE, a small shielded environment, 802.15.4, 868/434MHz radios.
More information at \url{https://doc.ilabt.imec.be/ilabt-documentation/}.
\\\hline

Fed4FIRE & 
European Commission sponsors the Fed4FIRE project, this is tooling and community for federation, not the testbed hardware itself.& 
See \url{https://www.fed4fire.eu/testbeds/} for the included testbeds. These testbeds are funded outside of the Fed4FIRE project.
\\\hline

GEANT Testbeds Service ("GTS")&
The GEANT Project (European Commission and 35+ European NRENs)&
GTS is a high performance, fully virtualized, automated SDX service that allows users to construct fully isolated and performance guaranteed network environments across Europe.   It provides VMs, Bare Metal Servers, SDN switching elements, and 10Gbps fully encapsulated NSI provisioned virtual circuits among these resources.  These environments are user defined and user controlled and can be set up literally in as little as a few seconds ready for use.   The GTS footprint includes London, Amsterdam, Paris, Prague, Hamburg, Madrid, Milan, and Bratislava.   \url{http://gts.geant.net} to register or for FFI.Other NRENs in Europe are deploying and will be available in early 2019:  NORDUnet (Copenhagen, Geneva, Washington DC, Miami), CESNet (Prague, Brno), DFN (Erlangen).  We are working to have other global pilots of this dynamic SDX capability in early 2019.
\\\hline

OF@TEIN+ SmartX Playground&
EU Asi@Connect&
OF@TEIN+ SmartX Playground is a miniaturized inter-national research testbed over Korea, Taiwan, and several South-East Asian countries. To assist in developing and operating IoT-SDN-Cloud functionalities, distributed and hyper-converged SmartX Boxes are composed into virtualized compute/storage/networking entities. Also from multiple (to-be-federated)  Playground Towers, all distributed SmartX Boxes are  provisioned remotely, inter-connected via SDN-coordinated overlay networking over TEIN and national R\&E networks, monitored together, and orchestrated for sustained playground operation. Recently OF@TEIN+ SmartX Playground is focusing on enabling the agile development for cloud-native (container-leveraged) IoT-Cloud services, while more automated, SDN-coordinated, and federated operation is pursued for the operation side.
\\\hline

EdgeNet&
U.S. National Science Foundation (NSF)&
EdgeNet is a worldwide, bottom-up, viral testbed that spreads by local action.  EdgeNet is a modern overlay network and distributed systems testbed in the spirit of PlanetLab, which was the most successful computer systems testbed in history.  EdgeNet has been informed by the advances of cloud computing and the successes of such distributed systems as PlanetLab, GENI, G-Lab, SAVI, and V-Node: a large number of small points-of-presence, designed for the deployment of highly distributed experiments and applications.  EdgeNet differs from its predecessors in two significant areas: first, it is a software-only infrastructure, where each worker node is designed to run part- or full-time on existing hardware at the local site; and, second, it uses modern, industry-standard software both as the node agent and the control framework.  The first innovation permits rapid and unlimited scaling: whereas GENI and PlanetLab required the installation and maintenance  of dedicated hardware at each site, EdgeNet requires only a software download, and a node can be added to the EdgeNet infrastructure in one minute. The second offers performance, maintenance, and training benefits; rather than maintaining bespoke kernels and control frameworks, and developing training materials on using the latter, we are able to ride the wave of open-source and industry development, and the plethora of industry and community tutorial materials developed for industry standard control frameworks.  The result is a global VM-hosted  Kubernetes cluster, where pods of Docker containers form the service instances at each point of presence.  EdgeNet is currently deployed at over 35 sites in the US, Canada, and the European Union, and PlanetLab Europe intends to convert all of its sites to EdgeNet.
\\\hline

Smart Gigabit Communities&
U.S. National Science Foundation (NSF)&
Twenty-four U.S. communities and two international communities are organized politically, academically, and corporately to support advanced smart and connected community applications and services that may depend upon advanced networking requirements.  While each community has unique characteristics and areas of application emphasis, there is a goal for all communities to have local gigabit interconnects that allow very low latency and very high bandwidth within-community applications and services  GENI and Edge-Net and sometimes community-specific resources are available for in-community cyberinfrastructure support.
\\\hline

Future Internet Brazilian Environment for Experimentation (FIBRE)&
Brazil's Ministry of Science, Technology, Innovation and Communication (MCTIC) and the National Education and Research Network (RNP)&
FIBRE is a large-scale research facility for experimentation on Future Internet technologies, composed by a federation of 16 resource centers (a.k.a. experimentation islands) hosted in universities and research centers in Brazil. FIBRE has its own Wide Area Network (WAN) backbone, built as a layer 2 SDN overlay network on top of RNP's backbone. Currently, FIBRE is a service operated by RNP and it is also used for teaching computer networking courses.
\\\hline

Chameleon (\url{www.chameleoncloud.org})&
U.S. National Science Foundation (NSF)&
Chameleon is a deeply programmable large-scale research testbed supporting Computer Science experiments in areas ranging from systems and networking to cloud computing and machine learning. Chameleon resources support experiments requiring hundreds of nodes, large-scale storage system in many configurations, or a diversity of hardware systems including GPUs, FPGAs, advanced memory systems, and virtualizable network switches. To date, Chameleon has been used by 2,700+ users working on 450+ Computer Science research and eduction projects. 
\\\hline

KREONET-S&
Ministry of Science and ICT, South Korea&
KREONET-S is a new network project to drive softwarization of KREONET Infrastructure with network virtualization, automation, and intelligence technology development. It is deployed as an SDN-WAN in eight locations now in Korea (5), the US (2), and China (1), supporting various advanced R\&E communities and users who want to experiment their innovative network technologies as well as to use KREONET-S for their very high performance data transfer in the automated and intelligent manner using virtually dedicated networking environment. KREONET-S is growing to contribute primarily to the new areas of hyper-convergence and data-centric ICT based on IoT, cloud, big data, supercomputing, and data-intensive science.
\\\hline

Grid'5000&
&
Grid'5000 is a large-scale and versatile testbed for experiment-driven
research in all areas of computer science, with a focus on parallel and
distributed computing including Cloud, HPC and Big Data. It is composed
of 8 sites located in France and Luxembourg, and 800 nodes featuring
various technologies: GPU, SSD, NVMe, 10G and 25G Ethernet, Infiniband, Omnipath.
It is highly reconfigurable, controllable and monitorable: experimenters
can use bare-metal deployment and network reconfiguration (for isolation
or building topologies), and monitor the network traffic and power
consumption of their nodes. \url{https://www.grid5000.fr/}
\\\hline

JGN&
National Institute of Information and Communications Technology
(NICT), Japan&
JGN is a high-speed network testbed operated by NICT, which provides a
wide-range of networking facilities to research and development
projects. It consists of domestic and international wide-area Ethernet
circuits, and is a member of the global R\&E (research and education)
network community with collaborations with SINET, Internet2, TransPAC,
Pacific Wave, SingAREN, CERNET, KOREN, KREONET, etc.
From a technical point of view, JGN networking services are based on
MPLS (ex. Ethernet over MPLS and VPLS) and router virtualization
(Juniper logical systems). JGN has the network slicing capability to
accommodate a wide-range of advanced networking experiments such as
uncompressed 8K video transmission and wide-area OpenFlow-based SDN
deployment. JGN also provides backbone networking facilities to other
NICT testbeds such as JOSE, StarBED, and RISE.
\\\hline

Japan-wide Orchestrated Smart/Sensor Environment (JOSE)&
National Institute of Information and Communications Technology
(NICT), Japan&
JOSE is an open
testbed operated by NICT that can accommodate multiple Internet of
Things (IoT) experiments consist of sensor networks, distributed
storage/computation resources and network resources. There are 5
distributed data centers in Japan (Yokosuka, Kanazawa, Kyoto, Tokyo,
Osaka) and more than 10,000 virtual machines (or containers) are
available for the experiments on these data centers.
The storage/computation/network resources are controlled and
virtualized by SDN and SDI functions in a centralized manner. Each
experimenter can deploy their processing modules on the distributed
servers in the dedicated network slice. The experimenters can use the
storage/computation resources physically located closer to the
experimenter's sensor network as an edge-cloud facility.
As one of the next steps of such testbed technologies, we are
developing functions based on two-layered (Platform as a Service layer
and Infrastructure as a Service layer) edge-cloud architecture for IoT
edge computing experiments.
\\\hline

Research Infrastructure for large-Scale Experiments (RISE)&
National Institute of Information and Communications Technology (NICT), Japan&
RISE is a
wide-area SDN testbed operated by NICT, which consists of SDN switches
and computer nodes at more than 10 locations including overseas
sites such as Seattle, Singapore, and Bangkok. RISE provides a
wide-area SDN environment to a user which can be fully controlled by
the user's controller. One of the major feature of RISE is its
flexibility in network topology. A RISE user can request the network
topology of her slice without considering the underlying physical
network configurations. Although the original objective of RISE was to
accommodate wide-area SDN-related experiments on JGN, its technical
coverage has been extended to IoT-related experiments with IoT
gateways. The IoT gateways provide a simple interconnection mechanism
between a RISE user slice and IoT devices, which decreases
environmental setting-up costs of IoT-related experiments.
\\\hline

StarBED&
National Institute of Information and Communications Technology (NICT), Japan&
StarBED is a general-purpose network emulation testbed operated by
NICT based on 1000+ baremetal PC servers and network switches located
in a single site (Ishikawa). Experimenters can install OSes and
application software to the nodes and can configure their complex
experimental topologies with VLANs. NICT has been developing a
software suite named SpringOS for the operation of StarBED such as
node power control, network configuration, and so forth.
Experimenters can conduct not only wired network experiments but also
wireless ones with NETorium that can emulate wireless communications
such as WiFi and BLE on the wired networks of StarBED.
Now, we are developing new mechanisms of wall-clock time coordination
between emulation environments and software simulation environments to
support experiments on StarBED in IoT technologies interacting with
physical phenomena.
\\\hline

cyber DEfense Technology Experimental Research Laboratory (DETERLab)&
U.S. Department of Homeland Security (DHS)&
DETERLab is a state-of-the-art scientific computing facility for cybersecurity researchers engaged in research, development, discovery, experimentation, and testing of innovative cybersecurity technology. DETERLab is a shared testbed providing a platform for research in cybersecurity and serving a broad user community, including academia, industry, and government. To date, DETERLab-based projects have included behavior analysis and defensive technologies including DDoS attacks, worm and botnet attacks, encryption, pattern detection, and intrusion-tolerant storage protocols.
\\\hline

Dispersed computing testbed (DcompTB)&
U.S. Defense Advanced Research Projects Agency (DARPA)&
DCompTB is a network testbed facility that allows researchers to rapidly design, deploy and execute complex experimental networked systems. Experiments involving hundreds of nodes can be materialized in minutes. The compute and network platforms provided by the testbed are open source hardware, providing maximum flexibility as an experimental platform. All nodes are have remotely accessible UART consoles for low-level systems development. The testbed provides advanced network modeling and emulation capabilities, allowing researchers to deploy systems in high operational-fidelity environments.
\\\hline

Advanced Instrumentation Measurement and Services (AMIS): Programmable Network Measurement for 100 Gbps Data Intensive Science Networks&
National Science Foundation, University of Massachusetts Lowell, University of Texas El Paso University of Massachusetts Boston, International Center for Advanced Internet Research, Northwestern University&
This testbed was designed and developed to experiment with techniques that can address the challenges of 100 Gbps (and beyond) network flows, as well as constantly changing traffic patterns, measurement targets and metrics, and policy differences across multiple network domains, especially for data intensive science applications. 
\\\hline

AutoGOLE / MEICAN Network Service Interface (NSI) Testbed&
Global Lambda Integrated Facility Community (GLIF), GLIF Open Lambda Exchanges (GOLEs), and multiple international R\&E networks.
\\\hline

BigData Express International 100 Gbps Testbed&
Department of Energy Office of Science, Fermi National Accelerator Laboratory, Oak Ridge National Laboratory, Energy Science Network (p), KISTI, KREONET, StarLight International/National Communications Exchnage Facility, Metropolitan Research and education Network, International Center for Advanced Internet Research, Northwestern University&
This testbed is suporting experimental research and demonstrations to develop services for large scale scientific data transfers, large capacity, high performance, E2E flows across thousands of miles and across multiple domains on 100 Gbps optical networks. Such transfer services support the collection, indexing, archived, sharing, and analysis of petabytes of data.
\\\hline

Ciena Environment for Network Innovations (CENI)&
Ciena&
The CENI testbed is a partnership project with the GENI testbed and its research communities and leverages GENI architecture and technologies. The CENI testbed was designed and developed to support dynamic SDN architecture and techniques on WAN L1 and L2 using 100 Gbps optical networks..
\\\hline

Ciena Research on Demand International 100 Gbps Network&
Ciena&
The Ciena international 100 Gbps testbed was developed in partnership with the International Center for Advanced Internet Research (iCAIR) at Northwestern University, the StarLight International/National Communications Exchange Facility Consortium, the Metropolitan Research and education Network (MREN), and CANARIE, the Canadian national R\&E network. This 100 Gbps testbed, which is based on thousands of miles of dedicated optical fiber, is being used to support SDN research experiments  realted to data intensive science flows over 100 Gbps optical networks.
\\\hline

Cisco Information Centric Networking Testbed (ICN)&
Cisco, I2, multiple partner universities&
This testbed was established to conduct experimental research with Information Centric Networking (ICN), a networking architecture and technology developed to provides an alternative to IP4/IPv6 protocols. Currebtly, a distributed research community is exploring the benefits of ICN, including on the US nation-wsie ICN Backbone Project, which has deploy and is operating the testbed to support research and experimentation.
\\\hline

Energy Science Network (ESnet) 100 Gbps National Testbed,&
ESnet&
The testbed provide research experimenters with access to high-speed hosts with 100G Ethernet NICs, fast RAID disk systems, running the latest Linux kernel, allows creation of point to point as well as multi-point circuits using SDN, options for virtual switches (OVS), virtual sites using VMs, measurements, protocol variants, and intelligent network architecture components.
\\\hline

GENI P4 Testbed&
National Science Foundation&
In partnership with eth GENI Program Office (GPO), te International Center for Advanced Internet Research at Northwestern University (iCAIR) is designing and developing a GENI P4 experiment/development environment, which will be implemented as an extension of the existing GENI SDX at the StarLight International/National Communications Exchange. This P4 testbed is being made available in stages. The testbed includes P4 Inventech switches based on the Barefoot Networks Tofino chip and P4 libraries in the Chameleon Cloud testbed.
\\\hline

International Global Environment for Network Innovations (iGENI)&
Originally, U.S. National Science Foundation (NSF) through the GENI Program Office (GPO), subsequently the International Center for Advanced Internet Research, Northwestern University and the Consortium for the StarLight International National Communications Exchange Facility, and the consortium for the Global Lambda Integrated Facility (GLIF)&
iGENI is an international federated Software Defined Networking (SDN) experimental testbed federated research testbed with sites in the US, Canada, Europe, South America and Asia. iGENI, which is connected to the US GENI testbed, supports various types of experiments based on SDN and Softwae Defined Exchange (SDXs).Ê
\\\hline

GENI Software Defined Exchange (G-SDX)&
Originally, U.S. National Science Foundation (NSF) through the GENI Program Office (GPO), subsequently the International Center for Advanced Internet Research, Northwestern University and the Consortium for the StarLight International National Communications Exchange Facility, and the consortium for the Global Lambda Integrated Facility (GLIF).&
The GENI SDX is an international federated Software Defined Exchange facility that enables experiments related to sliceable exchanges, programmable networking, network slicing, network stitching, programmable peering, federation, testbed federation, high performance transport protocols, specialized services for data intensive science, and other research topics.
\\\hline

International Petatrans Data Transfer Node (DTN) Testbed&
National Science Foundation, Global Lambda Integrated Facility (GLIF) community.&
This experimental research testbed is developing advanced network services for multidomain high performance data intensive ``petascale'' science, based on 100 Gbps Data Transfer Nodes (DTNs) integrated with SDN and dynamic network programming techniques.
\\\hline

Large Synoptic Survey Telescope Network Testbed&
National Science Foundation, LSST community, national Center for Supercomputing Applications (NCSA), Global Lambda Integrated Facility (GLIF) community&
This LSST networking testbed has been established to prepare for the production network services required by the LSST when it begins production, When production begins, the LSST will send hundreds of Gbps of data 7/24 from Chile to the National Center for Supercomputing Applications, where the data will be shaped into science products then distributed world-wide.
\\\hline

Metropolitan Research and Education Network (MREN) Optical Testbed(OMNInet)&
Metropolitan Research and Education Network (MREN)&
This testbed, consisting of dedicated (private) optical fiber and co-location sites in a metro area, was designed and developed to conduct research on dynamic provisioning of lightwaves, including dynamic lightpath switching) using DWDM and SDN techniques.
\\\hline

Large Hadron Collider Open Network Environment (LHCONE) Point-to-Point Service Testbed&
CERN, Large Hadron Collider networking community, SURFnet/NetherLight, StarLight International/National Communications Exchange Facility, Global Lambda Integrated Facility (GLIF) community, multiple other international R\&E open exchange points&
This international testbed is being used to design, develop, and operate a scalable point-to-point inter-domain service for data intensive science, with an initial focus on LHC networking. The service is based on L2 East-West protocols and architecture, but also is exploring L1 and L3 based services.
\\\hline

SCinet 400 Gbps Prototype Network&
Arista, SCinet, International Center for Advanced Internet Research, Northwestern University, California Institute of Technology, University of Southern California&
This is a multipoint 400 Gbps LAN implemented as a proof of concept network testbed  for research, experimentation and demonstrations at the SC18 ACM International High Performance Computing, Networking, Storage and Analytics Conference in Dallas Texas.
\\\hline

SCinet WAN Tbps Prototype Network&
SCinet, International Center for Advanced Internet Research, Northwestern University, NASA Goddard Space Fligt Center, Naval Research Lab&
This WAN optical based testbed is a multipoint Tbps national WAN testbed implemented as a proof of concept network testbed  for research, experimentation and demonstrations, including optical disaggregation,  at the SC16, SC17, and SC18 ACM International High Performance Computing, Networking, Storage and Analytics Conferencs, most recently at the SC18 conference in Dallas Texas..
\\\hline

SCinet DTN-as-a-Service Testbed Network&
SCinet, International Center for Advanced Internet Research, Northwestern University, Ecostream, DEll&
This SCinet DTN-as-a-Service testbed network implemented as a proof of concept network testbed  for research, experimentation and demonstrations at the Sc17 and SC18 ACM International High Performance Computing, Networking, Storage and Analytics Conference.
\\\hline

SD-WAN International Testbed&
Korea Institute of Science and Technology Information (KISTI), Daejeon, KISTI, KREONET2&
This SD-WAN international testbed is based on a 100 Gbps path between Daejeon, South Korea and the StarLigt International National Communications Exchange Facility in Chicago. The SD WAN is based on highly programmable components using ONOS as a foundation
\\\hline

FIT (Future Internet of Things)&
ANR (French government), European H2020 co-funding&
In order to efficiently address the challenges for designing advanced digital infrastructure, cornerstone of the digital transformation of our society,  it is crucial to equip researchers and practitioners with a wide range of scientific and experimental resources and tools. This is achieved by deploying and operating a large-scale platform providing access to cutting-edge technologies in wireless networking, IoT and Cloud.

FIT is a French test platform established in 2011, offering a wide choice of technologies through a single interface enabling access to a large number of configuration and monitoring tools. FIT enables experimentation across a broad range of subject and greatly reduces the cost required to establish and monitor an experiment on any of these subjects.

FIT constitutes thus an accelerator for the design of advanced technologies for the Future Internet. It allows the testing of protocols and applications covering a large set of needs, from the Internet of Things, to wireless radio networks and the cloud.

The federation architecture allows the combination of resources from multiple testbeds in a single experiment. This characteristic makes FIT unique at the international level.

Additional information is available from \url{https://fit-equipex.fr/}.
\\\hline
Citylab&
Flemish government in Belgium + European Commission Projects (Research and Infrastructure programmes)&
The CityLab testbed (part of City of Things programme) is a highly realistic, cross-technology testbed platform which validates key Smart Cities R\&D results and facilitates innovative Smart City experiments on top of a large-scale testbed environment. The key characteristics of the testbed are: (a) City deployment: CityLab covers a dense zone in the center of a major Belgian city; (b) Cross-technology: the CityLab nodes support major unlicensed wireless technologies such as WiFi, Bluetooth Low Energy, DASH7 and IEEE 802.15.4, supported by an own LoRa deployment; (c) Multi-purpose: experiments can cover either a dense and small location (e.g., a home) but can scale to the entire neighborhood; (d) Multi-level openness: we allow experimenters to interact with the testbed on three levels; (e) Communication-level: network researchers can deploy their own network protocols on top of the nodes to evaluate their solutions in a realistic city-wide network; (f) Data-level: the CityLab nodes can continuously monitor the city parameters through sensors which can be deployed depending on the experiment requirements; and (g) User-level: using a living lab approach we will engage with users, allowing them to provide their input on novel smart city applications.

The City of Things smart city testbed CityLab takes the next step, operating a Smart City IoT research testbed where industrial and academic research can be performed in a realistic, city-wide setting covering the complete eco-system for low-level wireless communications up to high-level living lab research.

Additional information is available from \url{http://www.citylab.tech}.
\\\hline
\end{mpsupertabular*}
\twocolumn

\bibliography{GEFI2018FinalReport}{}
\bibliographystyle{plain}

\end{document}